\documentclass[11pt,a4paper]{article}
\pdfoutput=1 
\usepackage{jheppub}
\usepackage[utf8]{inputenc}
\usepackage{amsmath,amssymb,amsthm}
\usepackage{physics,mathtools,tikz,float}
\usepackage{hyperref}
\usepackage{caption}
\usepackage{subcaption}

\makeatletter
\def\@fpheader{\vspace{0.1mm}}
\makeatother
\newtheorem{theorem}{Theorem}
\usepackage{colortbl}
\definecolor{ddred}{RGB}{198,31,31}
\definecolor{ddgreen}{RGB}{0,160,77}
\usepackage{hyperref}
\hypersetup{
    colorlinks=true,
    linkcolor=ddred,
    filecolor=blue,      
    urlcolor=ddred,
    citecolor=ddgreen}

\def\CG{\mathcal{G}}

\def\CL{\mathcal{L}}

\title{Bouncing geodesics, black hole singularities, and singularities of thermal correlators}

\author{Sašo Grozdanov,$^{a,b}$}
\author{Samuel Valach,$^a$}
\author{and Mile Vrbica$^{a}$\linebreak\vspace{-0.2cm}}

\affiliation[a]{Faculty of Mathematics and Physics, University of Ljubljana, SI-1000, Ljubljana, Slovenia}
\affiliation[b]{Higgs Centre for Theoretical Physics, University of Edinburgh, Edinburgh, EH8 9YL, Scotland}

\abstract{Bouncing geodesics have been used as valuable probes of black hole singularities. In the dual boundary theory, the presence of bouncing geodesics is encoded in the analytic structure of correlation functions. Thus, when their existence is related to the presence of a black hole singularity, this presents a practical holographic framework to analyse, diagnose, and classify spacetimes with curvature singularities. To make this intuition precise, we use the Hadamard theory of hyperbolic differential equations to prove that both bulk and boundary retarded propagators diverge whenever two points can be connected by a null geodesic. We clarify why this statement remains valid beyond the geodesic regime (for operators of any dimension) and examine how holographic renormalisation modifies the structure of the dual propagator. We also present a general characterisation of bouncing geodesics and the associated singularities in correlation functions for arbitrary spacetimes. Furthermore, we compare the analytic structure of the correlators in position and momentum space and discuss explicit examples. Finally, we demonstrate the validity and concrete limitations of the bouncing geodesic approach to the study of black hole singularities. In particular, we show an explicit example of a black hole in the self-dual linear axion model, which has a curvature singularity despite the absence of bouncing geodesics.}

\begin{document}

\maketitle


\section{Introduction and summary}
\label{s.intro}

The two most recognisably characteristic properties of spacetime geometries with a black hole are the event horizon and the singularity. At least semi-classically, the physics of event horizons is relatively well understood. Most notably, they are surfaces of empty space that radiate quanta of energy in the thermal spectrum with the so-called Hawking temperature. The understanding of the structure of spacetime singularities, however, remains one of the central challenges in theoretical physics. In classical General Relativity, curvature singularities signal the breakdown of the geometric description of spacetime and, moreover, no semi-classical description can be trusted. In principle, the gauge/gravity duality (holography) \cite{Maldacena:1997re, Gubser:1998bc, Witten:1998qj} is (tautologically) supposed to provide us {\em all possible} insights into the physics of singularities. However, extracting this information in a useful manner has proven very difficult even in the semi-classical large-$N$ limit of the duality. While the dual conformal field theory (CFT) imprint of an event horizon is clear -- the CFT behaves as if in a thermal state -- the CFT imprints of the black hole singularity on field theory observables, such as thermal correlation functions, are significantly more difficult to extract and analyse.

A fruitful line of research addressing this question has focused on the so-called \textit{bouncing geodesics} as probes of both the black hole interiors and spacetime singularities \cite{Fidkowski:2003nf,Festuccia:2005pi}. In asymptotically anti–de Sitter geometries, null limits of a space-like geodesic can traverse the bulk, ``reflect off'' high-curvature regions, and run to the boundary of the second exterior region of the (eternal) black hole. While such trajectories were known to leave certain distinct imprints in boundary correlation functions, their precise nature remained somewhat unclear. For example, as argued in the original papers \cite{Fidkowski:2003nf,Festuccia:2005pi}, these imprints are not visible on the physically relevant principal sheet of the Wightman function (where the function is known to be analytic) and a non-trivial analytic continuation is needed to reveal them. 

More recent developments have clarified that this seeming imprint of a singularity exists beyond the geodesic limit and that one can identify the corresponding singularities in the dual propagator for any finite conformal operator dimension (bulk field mass) \cite{Ceplak:2024bja}. Concretely, by resumming the boundary operator product expansion (OPE), it was found that there is a corresponding ``bouncing singularity'' in the stress-tensor sector of the dual Wightman function. Crucially, however, this is only one of the sectors that contribute to the full correlator, which, as discussed above, must {\em not} exhibit these singularities on the physical sheet. Various methods have since been developed to analyse these bouncing singularities from different points of view. They include the OPE position space approach \cite{Ceplak:2024bja,Valach:2025saf,Ceplak:2025dds,Araya:2026shz} and various momentum space approaches using the WKB approximation \cite{Afkhami-Jeddi:2025wra,Jia:2025jbi,AliAhmad:2026wem}, bulk phase-shifts \cite{Jia:2026pmv} and properties of asymptotic quasinormal modes (QNMs) \cite{Dodelson:2025jff}.

Despite this progress, it appears that several important questions about our ability to diagnose the singularities of black holes from the singularities of dual thermal CFT correlators remain unanswered:
\begin{enumerate}
    \item \textit{The bouncing singularities seen in correlators, corresponding to bouncing geodesics, exist outside the geodesic regime, for any $\Delta$. Why is this so?}
    \item \textit{Can these singularities be studied directly without the need for analytic continuations of correlators to ``non-physical'' sheets or by studying separate sectors of correlators?}
    \item \textit{Do bouncing geodesics and corresponding singularities exist in more general spacetimes, and can one extend their notion to non-holographic (AdS/CFT) setups?}
    \item \textit{Is the existence of bouncing geodesics and bouncing singularities in thermal CFT correlators equivalent to the existence of a dual black hole curvature singularity, and, thereby, are its signatures the signatures of black hole singularities?}
\end{enumerate}
In this paper, we aim to provide an answer to all four questions. As will become clear, the central ingredient, at least to shed light on the first three questions, will be a theorem by Hadamard discussed below in our Theorem~\ref{t.hadam}.

The idea is as follows: Firstly, as motivated by a recent WKB analysis\footnote{Importantly, unlike in the Wightman function, \cite{Afkhami-Jeddi:2025wra} found that the bouncing singularities in AdS-Schwarzschild spacetime should be present on the principal sheet of the retarded propagator.} \cite{Afkhami-Jeddi:2025wra} and their privileged role in (causal) linear response theory, we identify the boundary retarded thermal correlator as the cleanest and most suitable object for the study of singularities. Secondly, we note that in this case, one can describe the corresponding bulk solution using the Hadamard theory of hyperbolic differential equations \cite{hadamard1923lectures}. In particular, for scalar fields of any mass (and as we discuss below, also for other fields), the bulk-to-bulk retarded Green's function $\mathcal{G}(X,Y)$ between two bulk points $X$ and $Y$ can be expressed as \cite{friedlander1975wave}
\begin{equation}
    \mathcal{G}(X,Y)=W(X,Y)\frac{1}{\mathcal{L}^{D-2}},
\end{equation}
for any odd-dimensional spacetimes ($D$ is the number of spacetime dimensions), where $W(X,Y)$ is a smooth function and $\mathcal{L}$ is the geodesic distance between $X$ and $Y$. The result in even $D$ has an analogous singularity, but the correlator behaves as $\delta(\CL^2)$ rather than as $1/\CL^{D-2}$. As is usual with solutions to wave equations, the wavefront in even $D$ is sharp instead of having a tail. Taking the boundary limit to access the dual CFT correlator, this implies the following central statement:\\
\linebreak
\textit{In any holographic theory, a thermal retarded propagator has a singularity whenever the two points that it depends on can be connected by a null geodesic.}\\
\linebreak
Despite the fact that certain versions of this claim have been assumed in the literature in the past years, the Hadamard theorems make this statement rigorous. The theorem then helps with deriving the following answers to the above four questions:
\begin{enumerate}
    \item The singularities in the propagator appear independently of the mass of the bulk field, i.e., for any conformal dimension of the dual scalar operator. The statement of the null-connectability is a statement at the level of the locations rather than specific values of the fields.
    \item The Hadamard theorem provides a rigorous framework, in which the lightcone singularity, bouncing singularities or bulk-cone singularities \cite{Dodelson:2023nnr} are its immediate consequences. In this language, the singularities are present on the principal sheet of the full physical retarded propagator, unlike in the Wightman function or in the two-sided propagator. 
    \item Even when the notion of a boundary dual is not available or unknown (that is, e.g., in asymptotically-flat and de Sitter spacetimes), one still finds corresponding singularities in the bulk-to-bulk retarded Green's function. This makes the bouncing geodesics a meaningful probe of the curvature singularities even outside the realm of AdS/CFT.
    \item The existence of bouncing geodesics and their associated correlator singularities is {\em not}, in general, equivalent to the existence of a black hole singularity. This is because there exist black hole singularities that give a ``softer" behaviour to the geodesic equation potential at a genuine curvature singularity, which is insufficient to give rise to a bouncing geodesic. We show this with an explicit (counter)-example. 
\end{enumerate}

With the help of this framework, we provide general definitions of bouncing geodesics and bouncing singularities -- independent of the concrete spacetime or the existence of a dual -- and prove a general condition (a requirement) on spacetimes that admit bouncing geodesics. We find that any spacetime with a diverging blackening factor at the curvature singularity leads to the existence of a bouncing singularity. Importantly, therefore, using this intuition, we are able to present explicit examples of black hole spacetimes that have curvature singularities, but no bouncing geodesics, thereby answering Question 4. Another aspect of our work is a discussion from the point of view of momentum space, which follows the approach of \cite{Dodelson:2023vrw}. In particular, we discuss the singularities in the complex time at constant momentum, and argue that such singularities may be understood as ``candidate'' singularities of the final position space correlators. In this work, we refer to singularities, which are eliminated by the Fourier integration over spatial momentum, as the ``phantom singularities''. We discuss two explicit examples that include such cases in which the analytic structure of the Green's function may be analysed both at fixed momentum and fixed spatial separation.

The rest of this paper is organised as follows. In Section~\ref{s.hadamardthms}, we introduce the central Hadamard theorem for the retarded bulk-to-bulk Green's function and derive the corollary for the retarded boundary propagator. In Section~\ref{s.BG}, we define bouncing geodesics and bouncing singularities on a general Lorentzian manifold. We provide a criterion to analyse which spacetimes admit such geodesics, discuss examples and explain the important role of holographic renormalisation. In Section~\ref{s.momspBG}, we follow the approach of \cite{Dodelson:2023vrw} and discuss the bouncing singularities directly from the structure of QNMs. Within this framework, we study examples for which there exist no bouncing geodesic and show explicit examples of the analytic structure of Green's functions for both fixed momentum and spatial separation.  We conclude our discussion in Section~\ref{s.disc} and introduce potential future directions. In Appendix~\ref{a.exampleBB}, we demonstrate the validity of the Hadamard theorem on concrete examples. Appendix~\ref{a.spectraldensity} discusses technical details of thermal correlators, and in Appendix~\ref{a.fourik}, we provide details of the relevant Fourier integrals used in the main text.


\section{The Hadamard theory}\label{s.hadamardthms}

In this section, we build the general framework that we will be using throughout this work, state the Hadamard theorem and discuss its implications.

\subsection{Setup}\label{ss.HADdefs}

The main object that we investigate is the retarded thermal propagator (Green's function) of a scalar operator $\mathcal{O}$ with dimension $\Delta$, which is defined by
\begin{equation}
    G(x,x')\equiv i\theta(t-t')\frac1Z\Tr\left\{e^{-\beta H}\left[\mathcal{O}(t,\mathbf{x}),\mathcal{O}(t',\mathbf{x}')\right]\right\},
\end{equation}
where $\theta(t-t')$ is the Heaviside distribution, $\beta$ is the inverse temperature and $x=\{t,\mathbf{x}\}$, $x'=\{t',\mathbf{x}'\}$ are the positions of operator insertions in a $d$-dimensional (boundary) spacetime. 

In holography, the retarded propagator $G(x,y)$ is obtained as the boundary limit of the retarded bulk-to-bulk Green's function $\mathcal{G}(X,Y)$. Here, $X$ and $Y$ are positions of the field insertions in the dual $D=(d+1)$-dimensional bulk. Assume that we can separate the bulk coordinates as $X=\{r,x\}=\{r,t,\mathbf{x}\}$, where $x$ are the coordinates along the conformal boundary and $r$ is the ``radial'' bulk coordinate. The conformal boundary is located at $r\to\infty$. More precisely,
\begin{equation}\label{e.getG}
    G(x,y)=\lim_{r,r'\rightarrow\infty}(rr')^{\Delta}\,\mathcal{G}\left(\{r,x\},\{r',y\}\right),
\end{equation}
where holographic renormalisation ensures that the limits are finite.

From the bulk perspective, the retarded Green's function $\mathcal{G}(X,Y)$ can be defined for any bulk Lorentzian metric by
\begin{equation}
    \left(\Box_X-V(X)\right)\,\mathcal{G}(X,Y)=\frac{\delta(X-Y)}{\sqrt{-g}},\qq{such that}{\rm{supp}}\,(\mathcal{G}(X,Y))\subset J^+(Y),
\end{equation}
where $J^+(Y)$ is the set of all points that can be reached along future directed causal geodesics from $Y$. $V(X)$ is a smooth function of $X$,\footnote{$V(X)$ can also include terms of the form $v_a(X)\nabla^a_{\!\!X}$.} that, e.g., for a free scalar field simply reads $m^2$. Note that the notion of $\mathcal{G}(X,Y)$ exists even if the spacetime is not asymptotically AdS and there is no notion of a boundary theory. In the cases when the bulk metric corresponds to an asymptotically-AdS black hole, the condition ${\rm{supp}}\,(\mathcal{G}(X,Y))\subset J^+(Y)$ is equivalent to selecting the incoming wave solution at the black hole horizon, and one obtains the dual thermal retarded propagator via \eqref{e.getG}.

Finally, note that in some cases, translational symmetry allows us to express $G(x,y)=G(x-y)$. Then, it is convenient to set $y=0$ and think of $G$ as a function of a single coordinate.

\subsection{Position space singularities of retarded Green's functions}\label{ss.HADthms}

Having set the definitions, we can now investigate general analytic properties of these objects. For the bulk-to-bulk retarded Green's function $\mathcal{G}(X,Y)$ the analytic structure was analysed more than a century ago in the context of Hadamard's theory of linear hyperbolic differential equations \cite{hadamard1923lectures}. The resulting analytic behaviour can be summarized in the following central Hadamard theorem \cite{friedlander1975wave}:
\begin{theorem}[The Hadamard theorem]\label{t.hadam}
    For all $Y$ in the normal neighbourhood of $X$,\footnote{A normal neighbourhood of $X$ is a region where every point $X'$ can be connected to $X$ by a unique geodesic that lies entirely within that region.} the retarded Green's function $\mathcal{G}(X,Y)$ on any smooth $D$-dimensional spacetime takes the form
    \begin{equation}\label{e.had}
        \mathcal{G}(X,Y)=\begin{cases}W(X,Y)\frac{1}{\mathcal{L}^{D-2\phantom{|}}}\phantom{\Big|}&\text{for $D$-odd}\\
        U(X,Y)\delta(\mathcal{L}^2)+Q(X,Y)\phantom{\Big|}&\text{for $D$-even}\end{cases}
    \end{equation}
    where $W(X,Y)$, $U(X,Y)$ and $Q(X,Y)$ are smooth functions, $\delta$ is the Dirac distribution and $\mathcal{L}$ is the geodesic length between $X$ and $Y$.
\end{theorem}
We note that in the form that this theorem is stated, we are assuming that $D > 2$.\footnote{Let us mention that this theorem remains valid for $\mathcal{G}$ being a tensor of arbitrary rank.} A crucial consequence of this theorem is that the bulk Green's function has a singularity whenever one can connect points $X$ and $Y$ by a null geodesic. In fact, \textit{all} singularities of $\mathcal{G}(X,Y)$ originate from this null-connectability. Let us briefly mention that Theorem \ref{t.hadam} is ``local'' in a sense that $Y$ must lie in a normal neighbourhood of $X$. While such a neighbourhood may not exist globally, it is known \cite{garabedian1998partial,Kay:1996hj,ikawa2000hyperbolic} that even outside a normal neighbourhood, $\mathcal{G}(X,Y)$ continues to diverge if the two spacetime points can be connected by a null geodesic -- this is known as the ``propagation of singularities theorem'', see \cite{futurepaper} for the proof.\footnote{While the singularity is still present, the concrete form of $\mathcal{G}$ may differ from \eqref{e.had}.}

Using Theorem~\ref{t.hadam}, we can make a precise statement for the boundary propagator in AdS/CFT in the presence of a black hole:
\begin{theorem}\label{t.bdryhadam}
    In any holographic theory, the retarded propagator $G(x,y)$ has a singularity whenever $x$ and $y$ can be connected by a null geodesic.
\end{theorem}
This theorem follows from a combination of Theorem \ref{t.hadam} for $\mathcal{G}(X,Y)$, the relation between $\mathcal{G}(X,Y)$ and $G(x,y)$, the fact that the conformal rescaling in \eqref{e.getG} does not affect the structure of null-geodesics, and the fact that holographic renormalisation only eliminates the UV divergences, not affecting the locations of the singularities. Note that despite both $x$ and $y$ lie on the conformal boundary, in general, the geodesic between them passes through the bulk interior. Importantly, null geodesics that connect points $x$ and $y$, at which time and space are complexified are also genuine singularities of the retarded correlator. We will see examples of such geodesics in Section~\ref{s.momspBG} where we consider infinite lattices of singularities in complexified time.

Let us also make a few remarks:
\begin{itemize}
    \item The retarded propagator $G(x,y)$ will in general not satisfy the Hadamard form \eqref{e.had}. While the singularity will still exist at the same location, the power and form of the singularity may change -- we will discuss this in more detail in Sec.\ \ref{ss.BGRN} as an effect of holographic renormalisation.
    \item Note that the Theorems \ref{t.hadam} and \ref{t.bdryhadam} do not require the geodesic limit (i.e., a large probe mass regime). They apply to bulk fields and boundary operators for general probe mass $m$, resp., for general $\Delta$. Theorem \ref{t.bdryhadam} therefore makes it clear why certain ``geodesic singularities'' are present even for general $\Delta$ \cite{Ceplak:2024bja}.
    \item Theorems \ref{t.hadam} and \ref{t.bdryhadam} can be equivalently formulated for the advanced correlator. This captures for example the recently studied case of a complex bulk-cone geodesic \cite{Araya:2026shz} with space-like separated points on the boundary.
    \item Importantly, Theorem \ref{t.bdryhadam} is only valid for the \textit{retarded} (resp.\ \textit{advanced}) propagator; here, one finds singularities on the principal sheet, without any restrictions on their concrete kinematic regime. This is in sharp contrast to the Wightman function, which is analytic on the principal sheet and to observe any non-trivial singularities, one needs to either perform complicated analytic continuations or examine specific parts of the propagator (e.g.,\ the stress-tensor sector) \cite{Ceplak:2024bja,Afkhami-Jeddi:2025wra,Dodelson:2025jff}.
\end{itemize}

\subsection{Examples}

The Hadamard theorems presented above form a universal framework in which various phenomena studied in the literature in the recent years arise naturally. Let us mention a few examples here.
\begin{figure}[ht!]
\centering
\includegraphics[scale=1.00]{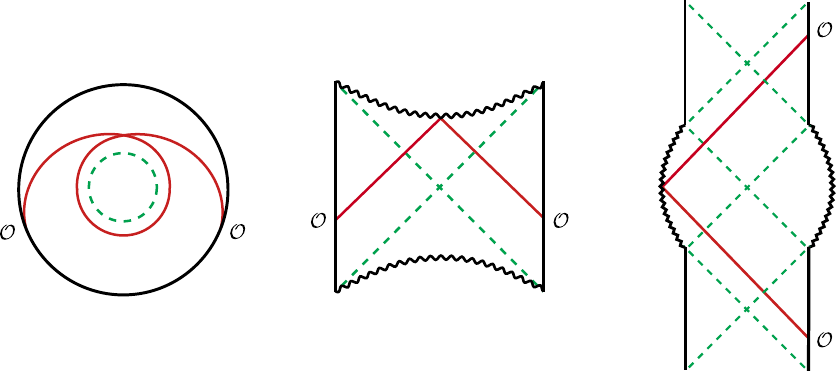}
\caption{Examples of null-geodesics (red) leading to singularities in the boundary retarded propagator. From left to right these are: geodesic propagating at the photon sphere of 1-sided AdS-Schwarzschild, geodesic bouncing off the space-like singularity in 2-sided AdS-black brane, geodesic bouncing off the time-like singularity in AdS-Reissner–Nordström black brane.}
\label{f.multipot}
\end{figure}

\paragraph{Lightcone singularity:} A trivial example of Theorem~\ref{t.bdryhadam} is the situation when one operator is placed on the (boundary) lightcone of the other operator. This is the case, e.g.,\ for AdS black brane for operators inserted on the conformal boundary in the one-sided picture. This is the so-called \textit{lightcone/OPE singularity}.

\paragraph{Bulk-cone singularities:} More non-trivial examples appear when the geodesic connecting the boundary operator insertions penetrates the bulk geometry. For the spherical black holes one finds a family of the so-called \textit{bulk-cone singularities} \cite{Hubeny:2006yu,Dodelson:2020lal,Dodelson:2023nnr,Auzzi:2025sep}. These correspond to points connectable by null geodesics propagating along the photon sphere, see Fig.\ \ref{f.multipot}. Generically, these are present at real Lorentzian times, although recently a complexified version of these geodesics were considered in \cite{Jia:2026pmv,Araya:2026shz}.

\paragraph{Bouncing singularities:} Another example are the \textit{bouncing singularities} that correspond to geodesics bouncing off a black hole curvature singularity \cite{Fidkowski:2003nf,Festuccia:2005pi,Ceplak:2024bja,Buric:2025anb,Buric:2025fye,Barrat:2025twb,Afkhami-Jeddi:2025wra,Ceplak:2025dds,Dodelson:2025jff,Jia:2025jbi,AliAhmad:2026wem,Jia:2026pmv,Araya:2026shz,Giombi:2026kdz}. These are less intuitive from the point of view of the 1-sided geometry, but they are natural probes of the black hole interior in the maximally extended spacetimes.\footnote{See Appendix~\ref{a.spectraldensity} for more detail on the 2-sided and 1-sided geometries and their interpretations.} They are generally found at complex times, e.g., for the AdS-black brane in $D=d+1$ dimensions and zero spatial displacement, they are located at
\begin{equation}\label{e.oldBS}
    t_*=\pm\frac{\beta e^{\mp\frac{i\pi}{d}}}{2\sin\frac{\pi}{d}}.
\end{equation}
The imaginary part of $t_*$ arises as a residue due to the horizon crossing in the complexified Schwarzschild coordinates. The analytic structure of the retarded boundary propagator thus provides important information about the presence and structure of the black hole interior in the dual geometry. For details regarding the analytic continuation of the retarded thermal Green's function, see Appendix~\ref{a.spectraldensity}.


\section{Bouncing geodesics: the position space perspective}\label{s.BG}

Let us now discuss the notion of bouncing geodesics and bouncing singularities in more detail, discuss their generalization, conditions for their existence and their role as a diagnostic of curvature singularities in AdS/CFT.

\subsection{Definitions and existence}\label{ss.GBdefs&gen}

Assume a spacetime with a curvature singularity, which is, without loss of generality, located at $r=0$. Then define bouncing geodesics and bouncing singularities as follows:
\begin{itemize}
    \item \textit{Bouncing geodesic is a null limit of a space-like or time-like bulk geodesic that approaches the curvature singularity from a finite distance $r_i$, comes infinitesimally close to it, and then moves finitely far away from it to some $r_f$ (i.e., it ``bounces off the curvature singularity'').}
    \item \textit{Bouncing singularity is the singularity of $\mathcal{G}(X,Y)$ (resp.,\ $G(x,y)$) corresponding to the points connectable by a bouncing geodesic via Thm \ref{t.hadam} (resp.,\ Thm \ref{t.bdryhadam}).}
\end{itemize}
Having the precise definitions, we may examine various properties of bouncing geodesics. 

We start by discussing the physical nature of the bouncing geodesics. One should have the following physical picture in mind: Spacetimes with a curvature singularity are geodesically incomplete, one expects that any time-like geodesic -- that corresponds to a massive physical probe -- passing through the location of the curvature singularity ($r=0$) has to stop there irrespectively of its energy, i.e., the spacetime potential has a well at $r=0$. In the language of space-like geodesics, the corresponding potential has therefore a bump at $r=0$, allowing for a space-like geodesic to come arbitrarily close to $r=0$ and escaping again to a finite distance $r_f$. Importantly, if the potential has an infinite well, then the space-like geodesic approaches $r=0$ in the limit $E\to\infty$; we show that in this limit the geodesic becomes null, thus satisfying the  above definition of a bouncing geodesic. This statement can be formalised as follows:
\paragraph{Claim:}\label{t.existenceBG}
    \textit{For any $D$-dimensional spacetime of the form
    \begin{equation}\label{e.genBH}
        \dd s^2=-f(r)\dd t^2+\frac{\dd r^2}{f(r)}+r^2\dd\mathbf{x}^{\phantom{.}2}\qq{\rm where}f(r)\xrightarrow[]{r\to0}\pm\frac{b}{r^a}\qq{\rm and}a,b>0,
    \end{equation}
    there exists a bouncing geodesic and a corresponding bouncing singularity. For the case of the minus (plus) sign $f(r)$, the bouncing geodesic is a limit of a space-like (time-like) geodesic bouncing off a space-like (time-like) singularity.}

\begin{figure}[t!]
\centering
\includegraphics[scale=1.20]{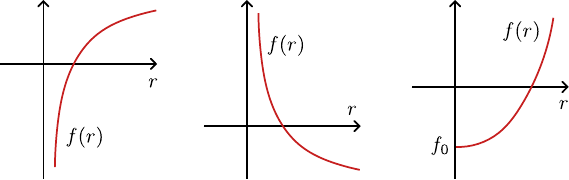}
\caption{Blackening factors $f(r)$ near $r=0$ for AdS-Schwarzschild (left), AdS-Reissner-Nordstr\"{o}m black hole (middle) and BTZ black hole (right). For the first two cases there exists a real geodesic bouncing off the space-like (resp.\ time-like) curvature singularity.}
\label{f.potentials}
\end{figure}

\paragraph{Proof:} To prove this, we set $\mathbf{x}=0$ and focus on the situation where $f(r)\to-b/r^a$.\footnote{For the opposite sign the proof follows the same steps.} For such spacetime, there exists a Killing vector $\partial_t$, which introduces a conserved quantity $E\equiv\dot{t}f(r)$. In addition, any (radial) geodesic has to satisfy
\begin{equation}\label{e.combinedeq}
    -f(r)\dot{t}^2+\frac{\dot{r}^2}{f(r)}=\epsilon\ \Longrightarrow E^2=\dot{r}^2-\epsilon f(r),        
\end{equation}
where $\epsilon$ is $0$, $+1$ or $-1$, for null, space-like and time-like cases, respectively. Note that one can treat $-\epsilon f(r)$ as a potential $V(r)$ probed by the energy $E^2$. 

The geodesic quantity we want to compute is the proper length of a radial geodesic $\mathcal{L}(E)$ starting from $r=r_i$ and ending at $r=r_f$ -- this is computed by an integral
\begin{equation}\label{e.lenghtgen}
    \mathcal{L}(E)=\pm\int_{r_i}^{r_f}\frac{\epsilon^2\dd r}{\sqrt{E^2+\epsilon f(r)}}=\pm2\int_{r_i}^{r_T}\frac{\epsilon^2\dd r}{\sqrt{E^2+\epsilon f(r)}},
\end{equation}
where in the second equality we assumed symmetric geodesics and introduced the turning point $r_T=r_T(E)$ defined as the largest real root of
\begin{equation}\label{e.solveforrt}
    E^2+\epsilon f(r)=0.
\end{equation}
Importantly, if the integral \eqref{e.lenghtgen} diverges for some choice of $r_i$ (typically for $r\rightarrow\infty$), one has to renormalise the length.

From the condition $\lim_{r\to0}f(r)=-\infty$, we immediately see that a space-like geodesic needs $E^2\to\infty$ in order to approach $r_T\to0$. Indeed, in this regime, one finds
\begin{equation}\label{e.rtaspartofprof}
    r_T(E)\approx b^{\frac1a}E^{-\frac2a}\to0,
\end{equation}
i.e., in this regime one discovers the required bouncing behaviour. We conclude the proof by showing that such space-like geodesic indeed becomes null in the limit $E\to \infty$, i.e., that $\mathcal{L}\xrightarrow[]{E\to\infty}0$.\footnote{Apart from the effects of renormalisation that we will discuss below. Note that case $a=1$ can be studied separately, yielding the same conclusions.}

Setting $r_i$ close to $r=0$ (but at a finite distance from it), we can always achieve that the integrand of \eqref{e.lenghtgen} is completely dominated by $1/E$ behaviour away from the actual curvature singularity. Therefore the only non-zero contribution in the large-$E$ limit can arise from the infinitesimal region near the black hole singularity. Here one needs to be careful, since in the denominator we have two competing effects: $E^2\to\infty$ and $f(r)\to-\infty$ that can, in principle mutually vanish. The indefinite integral \eqref{e.lenghtgen} can be solved close to the curvature singularity, yielding\footnote{One may be worried about the sequence of limits. However, here we follow the standard order: we first solve the integral (for small $r$-regime in this case), getting a family of space-like geodesics parametrized by $E$, and then send $E\to\infty$.}
\begin{equation}
    \pm2\int\frac{\dd r}{\sqrt{E^2+ f(r)}}\approx\pm\frac{2r \sqrt{E^2-b\,r^{-a }}}{E^2} \, _2F_1\left(1,\frac{1}{2}-\frac{1}{a };\frac{a -1}{a };\frac{b\,r^{-a
   }}{E^2}\right).
\end{equation}
Now, setting $r\to r_T(E)$ and then sending $E\to\infty$, we get\footnote{Here, we have used the fact that the contribution at $r=r_i$ vanishes in the large-energy limit.}
\begin{equation}
    \mathcal{L}(E)\xrightarrow[]{E^2\to\infty}E^{-\frac{a+2}{a}} \, _2F_1\left(\frac{1}{2},-\frac{1}{a};\frac{a-1}{a};b\right),
\end{equation}
which goes to 0 for any $a>0$. In other words, the geodesic we have now constructed satisfies the definition of a bouncing geodesic. Existence of the corresponding bouncing singularity is then a trivial consequence of Thm.\ \ref{t.bdryhadam}, which completes the proof.

Finally, let us make a few remarks:
\begin{itemize}
    \item The mentioned criteria are sufficient but not necessary. Even in spacetimes that do not satisfy \eqref{e.genBH}, bouncing geodesics may still exist; in such cases, they may appear as complex bouncing geodesics.
    \item Our proof does not require asymptotic AdS spacetime. In fact, one finds bouncing geodesics even in asymptotic dS \cite{Faruk:2023uzs,Faruk:2025bed} and other spacetimes \cite{futurepaper,futurepaper2}.
    \item An example of $f(r)\to+\frac{b}{r^a}$ is a charged AdS-black brane -- see the right diagram of Fig.\ \ref{f.multipot}. Indeed, as was shown in \cite{Ceplak:2025dds,Dodelson:2025jff,AliAhmad:2026wem}, there exists a corresponding singularity in the dual propagator. From the point of view of the Hadamard formalism, this is, again, a trivial consequence of Theorems \ref{t.hadam} and \ref{t.bdryhadam}.\footnote{Note that from the bulk Green's function point of view, the bouncing geodesic depicted by a red line in Fig.\ \ref{f.multipot} arises naturally as a null limit of a time-like geodesic. However, as shown in \cite{Ceplak:2025dds} and \cite{NejcesAxion}, from the boundary perspective, this geodesic is obtained as the limit of a space-like geodesic that becomes imaginary infinitesimally close to the curvature singularity.}
\end{itemize}

\subsection{Examples of absence}\label{ss.nono}

To understand the existence and limitations of the bouncing geodesics as probes of curvature singularities, we now examine explicit examples of black holes in which no bouncing geodesics exist.

\paragraph{The BTZ black hole:} Consider first a simple example of the neutral non-rotating BTZ black hole
\begin{equation}
   \dd s^2=-(r^2-r_0^2)\dd t^2+\frac{\dd r^2}{r^2-r_0^2}+r^2\dd x^2. 
\end{equation}
Importantly, since the corresponding potential well has a finite depth at $r=0$, i.e., $V(r=0)=f(r=0)=f_0$, no bouncing geodesics are guaranteed for this geometry. Indeed, it was shown in \cite{Fidkowski:2003nf} that in this geometry there are no bouncing geodesics attached to the conformal boundaries.\footnote{For more details see also \cite{Ceplak:2024bja}, where this case was examined using OPE techniques \cite{Fitzpatrick:2019zqz,Karlsson:2022osn,Huang:2022vet,Esper:2023jeq}, and also \cite{Amado:2008hw} where the BTZ was examined in the eikonal approximation.} For the BTZ black hole, the retarded boundary propagator is known explicitly and as we explain in the next section, there are no non-trivial singularities in the complex time plane, in accordance with the expectations from the Hadamard theorems. 

\begin{figure}[ht!]
\centering
\includegraphics[scale=1.1]{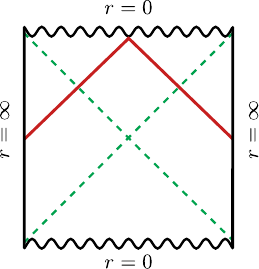}
\caption{Penrose diagram for the $D=3$ BTZ and the $D=4$ neutral self-dual axion black holes.}
\label{f.axionpenrose}
\end{figure}
Note that one may (incorrectly) assume there is a bouncing geodesic between the two asymptotic boundaries, see the red curve in Fig.\ \ref{f.axionpenrose}. Such a curve seems like a natural analogue of the black brane examples (cf., the middle diagram in Fig.\ \ref{f.multipot}) for a Penrose diagram where the curvature singularity is not bent -- the corresponding time would then be simply $i\beta/2$ corresponding to the crossing of the two horizons. However, such a curve is \textit{not} a geodesic, but rather a piecewise gluing of two null-geodesics. One can easily check that, unlike the black brane case, for the BTZ black hole, there is no space-like geodesic that would asymptote to the red curve in Fig.~\ref{f.axionpenrose}.

\paragraph{The self-dual linear axion model:} One may attempt to connect the absence of bouncing geodesics and bouncing singularities in the BTZ case to the fact that this geometry has no curvature singularity, only an orbifold singularity. However, consider now a different theory -- the holographic linear axion model in $D=4$ \cite{Andrade:2013gsa},\footnote{We do not consider the Maxwell field in this paper.} which is given by the action
\begin{equation}
    S=\int\dd^4x\sqrt{-g}\left(\mathcal{R}+6-\frac12\sum_{i=1}^2\partial^\mu\phi_i\partial_\mu\phi_i\right). \label{eq:axion}
\end{equation}
This theory has a simple black brane solution \cite{Davison:2014lua} (known as the model's self-dual limit, see also Refs.\ \cite{Grozdanov:2019uhi,Grozdanov:2025ner})
\begin{equation}\label{eq:axion-bh}
    \dd s^2=-(r^2-r_0^2)\dd t^2+\frac{\dd r^2}{r^2-r_0^2}+r^2(\dd x^2+\dd y^2),
\end{equation}
with the axion fields being $\phi_1=\sqrt{2}r_0 x$ and $\phi_2=\sqrt{2} r_0 y$. In spite of having a vanishing mass, this solution has an event horizon, and is dual to a thermal state with inverse temperature $\beta=2\pi/r_0$ and zero equilibrium energy density. 

Given that the blackening factor of metric \eqref{eq:axion-bh} is identical to that of the BTZ black hole, one can follow the same reasoning and computations to find that no bouncing geodesics exist in this geometry as well. In Section \ref{s.momspBG} we show that -- in accordance with the Hadamard theorems -- there are indeed no bouncing singularities in the retarded boundary propagator. Importantly, however, this model \emph{does} have a curvature singularity at $r=0$, since, for example, the Ricci scalar is given by
\begin{equation}
    \mathcal{R}=2\frac{r_0^2}{r^2}-12.
\end{equation}
This simple example therefore shows that the existence of a curvature singularity is, in general, \textit{not} equivalent to the existence of a bouncing geodesic or a bouncing singularity. 

\subsection{The role of renormalisation}\label{ss.BGRN}

Let us briefly discuss the subtleties and effects due to the boundary limit and holographic renormalisation. Concretely we will study what happens to the form of $\mathcal{L}$ and thus, indirectly, to the structure and type of the singularity in $\mathcal{G}$ and $G$.

We demonstrate this on a simple example -- black brane in AdS$_5$, for which $f(r)=r^2-\frac{\mu}{r^2}$, where the parameter $\mu$ is related to the mass of the black brane. In this case we can use the equations~\eqref{e.combinedeq}--\eqref{e.solveforrt} from the previous subsection; the indefinite integral for the proper length can be computed explicitly:
\begin{equation}\label{e.origino}
    I[r]\equiv\int\frac{\dd r}{\sqrt{E^2+f(r)}}=-\frac{r \sqrt{E^2-\frac{\mu }{r^2}+r^2} \ln \left(2 \sqrt{E^2 r^2-\mu +r^4}-E^2-2 r^2\right)}{2
   \sqrt{E^2 r^2-\mu +r^4}}.
\end{equation}
Importantly, between any two points $r_i$ and $r_f$ in the bulk interior (i.e., $r_i,r_f<\infty$), the actual geodesic length reads
\begin{equation}\label{e.clearpoly}
    \mathcal{L}(E)\equiv I[r_f]-I[r_i]\xrightarrow[]{E\to\infty}\frac{r_f-r_i}{E}+\mathcal{O}\Big(\frac{1}{E^3}\Big),
\end{equation}
which vanishes in the large energy limit (the regime of a bouncing geodesic), yielding the expected null-like behaviour $\mathcal{L}\to0$.

However, if one desires to anchor the geodesic at the asymptotic boundary $r\to\infty$, the geodesic length diverges, in accordance with the standard expectations from holography, and has to be renormalised. Holographic renormalisation dictates
\begin{equation}
    I[\infty]\equiv \lim_{r_{\rm max}\to\infty}\left(I[r_{\rm max}]-\ln r_{\rm max}\right),
\end{equation}
i.e., the geodesic length anchored at the conformal boundary takes the form $\mathcal{L}_{\rm RN}(E)\equiv I[r_f]-I[\infty]$. Remember that one always first computes a class of geodesics between anchoring points parametrized by $E$ and at the end takes any desired energy limit. Doing so, we find the following explicit expression:
\begin{equation}\label{e.wlogter}
    \begin{split}
        \mathcal{L}_{\rm RN}(E)&= \frac{1}{2} \ln \left(-\frac{E^4}{4}-\mu \right)-\frac{r_f \sqrt{E^2-\frac{\mu
        }{r_f^2}+r_f^2} \ln \left(2 \sqrt{E^2 r_f^2-\mu +r_f^4}-E^2-2r_f^2\right)}{2 \sqrt{E^2 r_f^2-\mu +r_f^4}}\\
        &\hspace{2cm}\xrightarrow[]{\phantom{k}E\to\infty\phantom{k}}\quad\ln \Big(\frac{E}{2}\Big)+\frac{r_f}{E}+\mathcal{O}\Big(\frac{1}{E^3}\Big),
    \end{split}
\end{equation}
where one can recognise the familiar logarithmic divergence in energy \cite{Fidkowski:2003nf} that, rewritten using $E(t)$, leads to a logarithmic divergence in $\mathcal{L}_{\rm RN}(t)$. Importantly, the above analysis clearly shows that the logarithmic divergence is \textit{always} present, and is a \textit{consequence of holographic renormalisation at the asymptotic infinity} rather than being a signature of the black hole singularity. One may choose $r_f$ outside the black hole horizon and still find the same logarithmic divergence.

Note that the large-energy limit of $I[r]$ always contains a term with $\ln(-E^2)$,
\begin{equation}
    \lim_{E\to\infty}I[r]\sim-\frac12\ln(-E^2),
\end{equation}
which, however, vanishes in the expression for proper length $\mathcal{L}(E)\equiv I[r_f]-I[r_i]$ for any $r_i,r_f<\infty$, yielding a simple polynomial behaviour in $1/E$ \eqref{e.clearpoly}. Only in the case when one (or both) of the anchoring points are at $r=\infty$, the subtraction of the logarithmic divergence happens only partially, yielding \eqref{e.wlogter}. In other words, the logarithmic divergence appears due to the non-commutativity of the limits
\begin{equation}
    \comm{\,\lim_{r_{\rm max}\to\infty}\,\,}{\,\,\lim_{E\to\infty}\,}\neq0.
\end{equation}

Importantly, note that the presence of the $\ln E$ term does not change the location of the bouncing singularity, which is still found at the time $t_*$ stated in Eq.~\eqref{e.oldBS}. It may, however, change the power and the type of the singularity -- indeed, even for finite $\Delta$, one finds the power of the singularity in $G$ to scale with $\Delta$ \cite{Ceplak:2024bja,Afkhami-Jeddi:2025wra}, as opposed to the $\Delta$-independent scaling in the Hadamard form \eqref{e.had} for $\mathcal{G}$. 

Finally, note that despite $\mathcal{L}\not\to0$, this geodesic is still null, as the $\ln(E)$ term should only be thought of as a relic of the renormalisation procedure.

\paragraph{Summary:} The process of holographic renormalisation eliminates the UV divergences but does not change the location of the singularities, which are given by the Hadamard theorems. However, since the boundary and the large-energy limits do not commute, the power of the singularity may change. The clear signature of such a curvature singularity is therefore only the location of the bouncing singularities due to the null-connectability of the operator insertions.


\section{Bouncing geodesics: the momentum space perspective}\label{s.momspBG}
We now discuss the manifestation of bouncing geodesics in momentum space boundary Green's functions, following \cite{Dodelson:2023vrw,Dodelson:2025jff,Afkhami-Jeddi:2025wra}. 

\subsection{Singularities from QNMs}

The momentum space retarded Green's function, defined as
\begin{equation}
    \hat G(\omega,\mathbf{k})=\int \dd t \dd \mathbf{x} \, G(t,\mathbf{x})\, e^{i\omega t-i \mathbf{k}\cdot \mathbf{x}}, \label{eq:xmas}
\end{equation}
is characterised (in the holographic theories of interest) by an infinite number of poles (quasinormal modes). In the simple cases of interest, they are organised into two asymptotic `Christmas tree' branches (see e.g.\ Ref.~\cite{Natario:2004jd}) $(\omega_n,-\omega_n^*)$ with\footnote{More branches are possible in the cases of finite charge density, finite coupling, or in the presence of higher-derivative corrections \cite{Grozdanov:2016vgg,Dodelson:2023vrw,Grozdanov:2018gfx,Jansen:2017oag}. Note also that subleading corrections to the asymptotic expansion \eqref{e.babac} depend on the boundary conditions of the bulk fields \cite{Grozdanov:2025ulc} and are therefore not expected to encode the information about the black hole singularity.}
\begin{equation}\label{e.babac}
    \omega_n(k) \sim \frac{n}{\zeta}+\ldots,\qquad n=0,1,\ldots,
\end{equation}
where $\zeta$ is a complex parameter, and ellipses denote the terms subleading in $n$. The imaginary part of $\zeta$ is fixed to \cite{Dodelson:2023vrw,Grozdanov:2025ulc}
\begin{equation}
\label{eq:imd}
    \Im \zeta =\frac{\beta}{4\pi}.
\end{equation}
It was conjectured \cite{Festuccia:2005pi,festucciathesis,Festuccia:2008zx,Dodelson:2023vrw} that, in the two-branched case discussed here, $\Re \zeta \neq 0$ indicates the presence of a curvature singularity, whereas $\Re \zeta = 0$ indicates its absence. This may be phrased as the behaviour of the correlator as $\omega \rightarrow i\infty^+$ \cite{Festuccia:2005pi} or as a set of sum rules \cite{Dodelson:2023vrw}. In Section \ref{s.phantom}, we provide an explicit counterexample for which $\Re \zeta = 0$ in spite of the presence of a curvature singularity, which may be connected to the absence of bouncing singularities in that model.

\begin{figure}
     \centering
     \begin{subfigure}[b]{0.495\textwidth}
         \centering
         \includegraphics[scale=1]{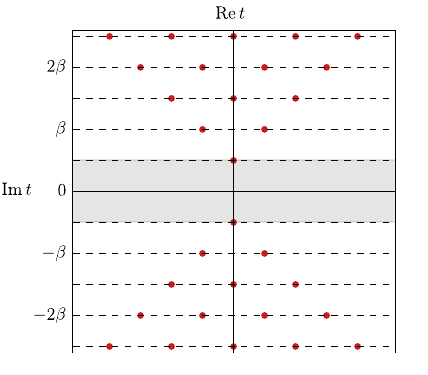}
     \end{subfigure}
     \hfill
     \begin{subfigure}[b]{0.495\textwidth}
         \centering
         \includegraphics[scale=1]{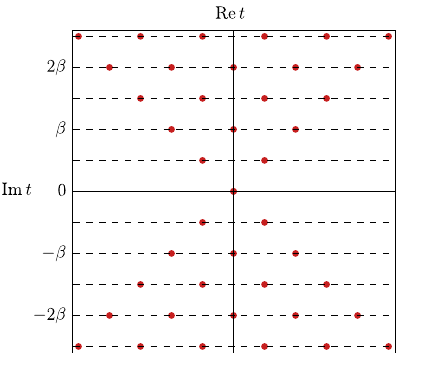}
     \end{subfigure}
        \caption{The singularities in the complex $t$ plane at fixed spatial momentum $\mathbf k$ for the two-sided correlator (left) and the retarded correlator (right). The former is characterised by a strip of analyticity for $-\frac{\beta}{2}<\Im t<\frac{\beta}{2}$.}
        \label{fig:open-tree}
\end{figure}

It was shown in Ref.\ \cite{Dodelson:2025jff}
by using transseries methods that such a configuration of poles gives rise to a lattice of singularities of the two-sided correlator $G_{12}$ in $(t,\mathbf{k})$ space (see Appendix~\ref{a.spectraldensity} for definitions) at points (see Fig.\ \ref{fig:open-tree}):
\begin{equation}
    \text{singularities of } G_{12}(t,\mathbf k): \qquad t=\pm\qty(i\frac{\beta}{2}+t_{mn}),\qquad n,m=0,1,\ldots,
\end{equation}
where we have introduced
\begin{equation}
    t_{mn}\equiv 2\pi \qty(\zeta n-\zeta^* m). 
\end{equation}
All singularities therefore lie at equally-spaced parallel lines to the real line, namely, at
\begin{equation}
    \Im t_{mn}=\frac{\beta}{2}\qty(n+m),
\end{equation}
together with their mirror images reflected across the real axis.

The retarded correlator is a linear combination of $G_{12}(t+i\beta/2)$ and $G_{12}(t-i\beta/2)$ (see Appendix~\ref{a.spectraldensity}). The singularities of the retarded correlator $G$ in $(t,\mathbf{k})$ space can therefore appear at (see Fig.\ \ref{fig:open-tree}):
\begin{equation}
    \text{singularities of } G(t,\mathbf{k}): \qquad t=\pm t_{mn},\qquad n,m=0,1,\ldots.
\end{equation}
In general, the singularities of the two terms are not expected to cancel. Note also that the parameter $\zeta$ is assumed to be $\mathbf{k}$-independent, and thus the position of the singularities in complex $t$ plane is $\mathbf{k}$-independent as well. This is true in all cases known to us.

\subsection{Phantom singularities and \texorpdfstring{$\mathbf{k}$}{k} integrals}
\label{s.phantom}
The lattice of singularities described above should be understood as existing at a constant spatial momentum $\mathbf k$, at least in systems with spatial dimensions. The boundary imprint of the bouncing singularity is, however, most easily understood in the standard position space $(t,\mathbf x)$ at $\mathbf x = 0$. It is therefore important to understand to what extent does the $(t,\mathbf k)$ analysis translate to results in the $(t,\mathbf x=0)$ space. Here, we propose that the singularities of the retarded (two-sided) correlator at $t=\pm t_{mn}$ ($t=\pm \qty[i\beta/2+t_{mn}]$) and fixed $\mathbf k$ should be understood as \emph{candidates} for the singularities at $\mathbf x=0$. This is due to the possibility of the Fourier integral over $\mathbf k$ resulting in non-trivial cancellations. For example, while the singularity corresponding to the bouncing geodesic \eqref{e.oldBS} is expected to be present both in the $(t,\mathbf{x})$ and the $(t,\mathbf{k})$ space, this is not expected to hold for any singularity in the complex $t$ plane. In this section, we provide tractable examples of such behaviour. The nature of singularities for $\mathbf x\neq 0$ is more complicated \cite{Araya:2026shz}, and we will not discuss their signatures in the $(t,\mathbf k)$ space here.

To understand what can happen with the singularities when performing the Fourier integral over momentum, we consider the two examples from Section \ref{ss.nono} -- the BTZ black hole and the self-dual linear axion model -- for which explicit expressions are available. Recall that in both theories, there is no singularity bending in the Penrose diagram (see Fig.\ \ref{f.axionpenrose}), so while there is also no curvature singularity in the BTZ black hole (only an orbifold singularity), the self-dual linear axion black hole does exhibit a genuine curvature singularity. By analysing the retarded boundary propagators (that can be computed in a closed form), we indeed confirm the existence of only the trivial lightcone singularities (and no bouncing singularities) in the complex time plane of $G(t,\mathbf{x}=0)$. This therefore allows us to see explicitly that not all of the singularities from the $(t,\mathbf k)$ space are translated into (branch point) singularities in the $(t,\mathbf x)$ space. While this has not been established here, it is possible that this property may hold even in certain more complicated models with a bouncing singularity \cite{Dodelson:2025jff}.

The special property of both of the examples considered in this paper is that the boundary retarded correlator takes the following form:
\begin{equation}
\label{eq:retarded}
    \hat G(\omega,\mathbf{k}) \propto \frac{\Gamma\qty(\xi(\mathbf{k})-i\frac{\beta\omega}{4\pi})\Gamma\qty(\xi^*(\mathbf{k})-i\frac{\beta\omega}{4\pi})}{\Gamma\qty(\xi(\mathbf{k})+\nu-i\frac{\beta\omega}{4\pi})\Gamma\qty(\xi^*(\mathbf{k})+\nu +i\frac{\beta\omega}{4\pi})},
\end{equation}
for some complex $\xi(\mathbf{k})$ and some real $\nu$, with $k$ being the magnitude of either one- or two-dimensional momentum. They both correspond to a case where $\Re\zeta = 0$, meaning that the asymptotic branches of poles in the complex $\omega$ plane are parallel to the imaginary axis. 

We first consider the singularities in the $(t,\mathbf{k})$ space. The Fourier integral is a Mellin-Barnes-type integral that appears in the fundamental integral representation of the Meijer G-function,
\begin{align}
    \widetilde G(t,\mathbf{k})\equiv \frac{1}{2\pi}\int \dd \omega \,\hat G(\omega,\mathbf{k}) \, e^{-i\omega t}\propto {\rm{G}}_{22}^{20}\qty(\mqty{\xi+\nu & \xi^*+\nu\\\xi & \xi^*};~e^{-\frac{4\pi t}{\beta}}),
\end{align}
which can, in turn, be expressed with the hypergeometric $_{2}F_1$ function \cite{erdelyi}:
\begin{equation}
    \widetilde G(t,\mathbf{k})\propto \theta(t) \qty[A(\xi,\nu)e^{-\frac{4\pi\xi t}{\beta}} {}_2F_1\qty(1-\nu,1-\nu+2i \Im \xi,1+2i\Im \xi;e^{-\frac{4\pi t}{\beta}})\\ +(\xi\rightarrow -\xi)],   \label{eq:hypergauss}
\end{equation}
where
\begin{equation}
    A(\xi,\nu)=\frac{\Gamma(-2 i \Im \xi)\, }{\Gamma(\nu)\Gamma(\nu-2i\Im\xi)\Gamma(1+2i \Im \xi)}.
\end{equation}
To proceed, we define
\begin{equation}
    t_n\equiv i\frac{\beta}{2}n,\label{eq:batsumara}
\end{equation}
for integer $n$. Since ${}_2 F_1(a,b,c;z)$ is singular at $z=1$, we get singularities of $\widetilde G(t,\mathbf{k})$ at $t_n$:\footnote{In the case of half-integer $\nu$, the singularity is logarithmic.}
\begin{equation}
    \widetilde G(t\rightarrow t_n,\mathbf{k})\sim (t-t_n)^{2\nu-1}, \qquad n=0,\pm1,\pm2,\ldots. \label{eq:Gret-candidates}
\end{equation}
By similar arguments, the two-sided correlator in $(t,\mathbf{k})$ space has singularities at
\begin{equation}
    \widetilde G_{12}(t\rightarrow t_n,\mathbf{k})\sim (t-t_n)^{2\nu-1}, \qquad n=\pm1,\pm2,\ldots. \label{eq:G12-candidates}
\end{equation}
This agrees with the analysis of Ref.\ \cite{Dodelson:2025jff} and ensures that there are no cancellations between singularities on the imaginary axis in the $(t,\mathbf k)$ space.

\begin{figure}
     \centering
     \begin{subfigure}[b]{0.3\textwidth}
         \centering
         \includegraphics[scale=1]{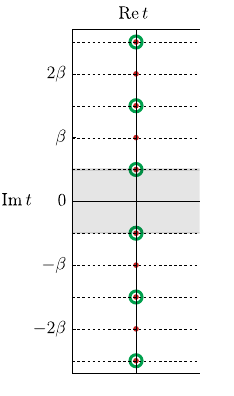}
         \label{fig:y equals x}
     \end{subfigure}
     \hspace{1cm}
     \begin{subfigure}[b]{0.3\textwidth}
         \centering
         \includegraphics[scale=1]{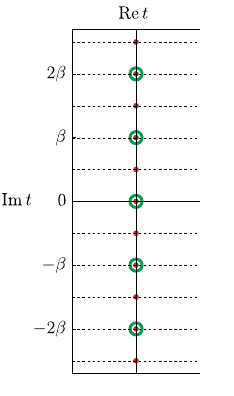}
         \label{fig:three sin x}
     \end{subfigure}
        \caption{The singularities in the complex $t$ plane in both the BTZ and self-dual linear axion cases at fixed $\mathbf{k}$ (red dots) and at $x=0$ (green circles) for the two-sided correlator (left) and the retarded correlator (right). Only a subset of the singularities survives the Fourier integration over $\mathbf{k}$.}
        \label{fig:closed-tree}
\end{figure}

Interestingly, a subset of these singularities does not survive the Fourier integral over $\mathbf{k}$. Constraining ourselves to the $\mathbf{x}=0$ case, the BTZ black hole and the axion model both exhibit the same singularity structure (see below). In the case of the retarded correlator, the singularities for \emph{odd} $n$ are not present in the $(t,\mathbf{x}=0)$ case, and only the even $n$ singularities remain (see Fig.\ \ref{fig:closed-tree}):
\begin{equation}
    \text{singularities of } G(t,\mathbf{x}=0): \qquad t=t_{2n},\qquad n=0,\pm 1,\pm 2,\ldots. \label{eq:Gret-true}
\end{equation}
Similarly, the even $n$ singularities are not present in the $(t,\mathbf{x}=0)$ case of the two-sided correlator, and only the odd $n$ singularities remain (see Fig.\ \ref{fig:closed-tree}):
\begin{equation}
    \text{singularities of } G_{12}(t,\mathbf{x}=0): \qquad t= t_{2n+1},\qquad n=0,\pm 1,\pm 2\ldots. \label{eq:G12-true}
\end{equation}
Eqs.\ \eqref{eq:Gret-true} and \eqref{eq:G12-true} should be contrasted with their $\mathbf{k}$-space counterparts \eqref{eq:Gret-candidates} and \eqref{eq:G12-candidates}.


\paragraph{The BTZ black hole with a scalar field:}\label{ss.BTZ}

The retarded Green's function for a scalar primary operator with dimension $\Delta$ dual to a minimally coupled scalar field on the background of a BTZ black hole takes the form of Eq.\ \eqref{eq:retarded} with
\begin{equation}
    \xi(k)= \frac{\Delta}{2}+i\frac{\beta}{4\pi}k, \qquad \nu=1-\Delta,
\end{equation}
where $k$ is the one-dimensional spatial momentum. The representation of the Green's function in $(t,x)$ space is completely determined by conformal invariance \cite{DiFrancesco:1997nk} (see \cite{Son:2002sd} and Appendix~\ref{app:btz} for a Fourier transform):
\begin{equation}
    G(t,x) \propto  \theta(t)\theta(t^2-x^2)\frac{\sin(\pi\Delta) }{\qty[\sinh\qty(\frac{\pi}{\beta}(t-x))\sinh\qty(\frac{\pi}{\beta} (t+x))]^\Delta}.
\end{equation}
The only singularities of this Green's function are at (see Eq.~\eqref{eq:batsumara})
\begin{equation}
    t=\pm x + t_{2n}, \qquad n=0,\pm1,\pm2,\ldots,
\end{equation}
which are the \emph{lightcone} singularities. At $x=0$, these correspond to the even candidates in \eqref{eq:Gret-candidates}, i.e., $t=t_{2n}$. There are no singularities at $t=t_{2n+1}$. The two-sided correlator, which may also be evaluated explicitly
\begin{equation}
    G_{12}(t,x)\propto  \frac{1}{\qty[\cosh\qty(\frac{\pi}{\beta}(t+x))\cosh\qty(\frac{\pi}{\beta}(t-x))]^\Delta},
\end{equation}
has singularities at
\begin{equation}
    t=\pm x + t_{2n+1}, \qquad  n=0,\pm1,\pm2,\ldots,
\end{equation}
which, at $x=0$, amounts to the set of singularities at $t=t_{2n+1}$, i.e., the odd candidates in \eqref{eq:G12-candidates}. 

Interestingly, the candidate singularities in the $(t,k)$ space from Eqs.\ \eqref{eq:Gret-candidates} and \eqref{eq:G12-candidates} fail to coincide with any singularities in the $(t,x)$ space for $x\neq 0$, where it becomes transparent that these singularities are of the lightcone type.

\paragraph{The self-dual linear axion model:}\label{ss.axion}
The retarded correlation function dual to the massless scalar in the self-dual linear axion model has the form of Eq.\ \eqref{eq:retarded}, with the following parameters \cite{Davison:2014lua}:\footnote{The fact that the self-dual linear axion model has a simple spectrum of equidistant poles parallel to the imaginary axis is interesting in and of itself. Here, we do not address the deeper reasons behind this structure and take the solution at face value.}
\begin{equation}
    \xi(k)=\frac{5}{4}+i\frac{\beta}{4\pi}p(k), \qquad \nu=-\frac{3}{2},
\end{equation}
where
\begin{equation}
    p(k)=\sqrt{\frac{7\pi^2}{\beta^2}+k_x^2+k_y^2},
\end{equation}
with $k_{x,y}$ being the components of the spatial momentum of the three-dimensional boundary theory. 

In this case, the Fourier transform cannot be performed explicitly. One can, however, express the $x=y=0$ value of the two-sided correlator with an integral (see Appendix~\eqref{app:axion} for the derivation):
\begin{equation}
    G_{12}(t,x=y=0)\propto \int_{-\infty}^\infty \dd s \frac{\sinh s}{s}\frac{\cos\frac{\sqrt{7}}{2}s}{\qty[\cosh\frac{2\pi t}{\beta}+\cosh{s}]^{\frac{7}{2}}}.
\end{equation}
The form of this integral now allows us to study its behaviour for complex $t$. More specifically, it is straightforward to conclude that the integral converges absolutely for $t=t_{2n}$, therefore the value of the two-sided correlator is strictly finite at that point. When $\Im t = t_{2n+1}$, the integral fails to be convergent. When $\Re t \neq 0$, this can be circumvented by an appropriate choice of the integration contour, which amounts to the choice of sheet, indicating the presence of a branch cut. The branch points are at $t=t_{2n+1}$, which is precisely the expected location of the singularities \eqref{eq:G12-true}. Consequently, the retarded correlator is strictly finite when $t=t_{2n+1}$ and has branch point singularities when $t=t_{2n}$, in accordance with Eq.\ \eqref{eq:Gret-true}. This is precisely the same as in the BTZ case, where only half of the candidates in the $(t,\mathbf{k})$ space end up as genuine position space singularities at $(t,x=y=0)$. 

Note that for both the BTZ black hole and the self-dual axion model, the phantom singularities are located at $i(\beta/2+2\mathbb{N}_0)$, which precisely corresponds to a piecewise null geodesic depicted by red curve in Fig.~\ref{f.axionpenrose}, and its natural ``multi-bounce'' generalizations. As we discussed in Section~\ref{ss.nono}, these piecewise null curves are \textit{not} geodesics, and hence -- following the Hadamard theorems -- cannot lead to singularities of the retarded boundary propagator. It would be interesting to understand whether the phantom points always arise as fictitious singularities corresponding to piecewise null geodesics ``bouncing off'' the $r=0$ surface.

\subsection{Lattice of singularities from the Hadamard theorem}

As explained above, for the BTZ black hole and the self-dual axion model, we only find the lightcone singularity and its KMS reflections in the retarded boundary correlator. Despite the fact that these are \textit{not} bouncing singularities, we can nevertheless find a geodesic ``origin'' of these singularities. This is in accordance with the Theorems \ref{t.hadam} and \ref{t.bdryhadam}. 

The lightcone singularity from the definition corresponds to the boundary points connected by a null line (on the boundary), which immediately leads to the corresponding singularity at $(t=0,\,\mathbf{x}=0)$. Let us now show that, in fact, all KMS reflections of the lightcone singularity  (the so-called \textit{KMS singularities}\footnote{More precisely -- since there is no KMS symmetry for the retarded propagator -- the term ``KMS singularities'' refers to the standard KMS singularities of $G_{12}$ once Eq.~\ref{appeq:rho-twosided} is applied. We assume without proof that none of these branch point singularities cancel out.}) correspond to null-geodesics, i.e., that
\begin{equation}
    {\rm{KMS\,\,\, singularities}}\,\,\,\,\,\longleftrightarrow\,\,\,\,\,\mathcal{L}=0.
\end{equation}

We start by performing a Wick rotation to the Euclidean time $\tau=it$. The metric then reads
\begin{equation}
    \dd s^2=f(r)\dd\tau^2+\frac{\dd r^2}{f(r)}+r^2\dd\mathbf{x}^2, \qquad f(r)=r^2-1,
\end{equation}
where we set $r_0=1$ for simplicity. The Penrose diagram for the Euclidean BTZ black hole (and the self-dual axion black hole) is the standard Euclidean circle (i.e., a projection of the Euclidean cigar into a plane). Set $\mathbf{x}=0$ and define the Euclidean energy along the geodesic as $\tilde{E}^2=\left(\dv{\tau}{s}\right)^2f(r)^2$. The proper length and proper time can then be computed via integrals
\begin{align}
    \mathcal{L}(\tilde{E})&=2\int_{r_i}^{r_T}\!\!\frac{\dd r}{\sqrt{f(r)-\tilde{E}^2}}=2 \ln\left(r_i+\sqrt{-1+r_i^2-\tilde{E}}\right)-\ln\left(1+\tilde{E}\right),\label{e.proplenin}\\
    \tau(\tilde{E})&=2\int_{r_i}^{r_T}\!\!\frac{\tilde{E}\dd r}{f(r)\sqrt{f(r)-\tilde{E}^2}}=-2\,\text{arctan}\left(\frac{r_i^2-1-r_i \sqrt{-1+r_i^2-\tilde{E}^2}}{\tilde{E}}\right)\!+\!2\,\text{arctan}(\tilde{E}),
\end{align}
where $r_T=\sqrt{1+\tilde{E}^2}$ is the turning point and we assumed a symmetric geodesic starting from $r=r_i$.
The expression for the proper time can be inverted analytically, yielding
\begin{equation}
    \tilde{E}^2=\frac{2 \left(r_i^2-1\right) \cos\left(\frac{\tau}{2}\right)^2}{1+r_i^2+\left(1-r_i^2\right)\cos(\tau)}.
\end{equation}
Using this expression for $\tilde{E}$ in Eq.\ \eqref{e.proplenin}, we get
\begin{equation}
    \mathcal{L}(\tau)=2 \ln\left(1+\sqrt{1-\frac{2}{1+\cos(\tau)+r_i^2-\cos(\tau)r_i^2}}\right)-\ln\left(\frac{2}{1+\cos(\tau)+r_i^2-\cos(\tau ) r_i^2}\right).
\end{equation}
Importantly, evaluating this function at multiples of $\beta/2$,\footnote{In the units where $r_0=1$, one has $\beta=2\pi$.} we find\footnote{For $r_i=\infty$, one needs to renormalise the proper length. This gives $\mathcal{L}=\ln4$ for any $\tau=\beta(\frac12+n)$, while for the times $\tau=\beta n$, the proper length logarithmically diverges, in accordance with the discussion in Section \ref{ss.BGRN}.}
\begin{equation}
    \mathcal{L}=\begin{cases}
        0\hspace{3.13cm}\qq{for}\tau=\beta n,\,\,\,n\in\mathbb{Z},\\
        2 \ln\left(r_i+\sqrt{r_i^2-1}\right)\qq{for}\tau=\beta(\frac12+n),\,\,\,n\in\mathbb{Z}.
    \end{cases}
\end{equation}
Hence, we see that the null-connected points precisely correspond to the locations of the KMS singularities \eqref{eq:Gret-true}, while for the ``phantom singularities'', one finds a non-zero $\mathcal{L}$. 

These results show that the Theorems \ref{t.hadam} and \ref{t.bdryhadam} indeed reproduce all singularities of the retarded correlator, even in cases where these geodesics and singularities are complex. 


\section{Discussion}\label{s.disc} 

In this work, we discussed a number of issues that have emerged in recent investigations of bouncing geodesics, singularities of dual holographic thermal correlation functions, and their relation to potential signatures of black hole singularities. Our work first used the results from the Hadamard theory of hyperbolic differential equations, which establishes a clear one-to-one relation between the existence of null geodesics and position space singularities of the retarded boundary correlator. Then we provided general definitions of bouncing geodesics and proved a general condition on spacetimes that admit bouncing
geodesics. We discussed the role of holographic renormalisation and analysed two simple examples -- the BTZ black hole and the self-dual axion model -- both of which exhibit no bouncing geodesics. We also derived the bouncing singularities from the momentum space perspective and examined the explicit structure of the retarded propagators for the BTZ and the self-dual axion black holes.

The most important question that remains open is what precisely are the unique and unambiguous signatures of the black hole singularities. While bouncing geodesics and their associated Green's function singularities are certainly correlated to the existence of a black hole singularity in many known examples, as we saw, the example of the self-dual linear axion black hole shows that this is not always the case. While this black hole is special in the sense that its mass vanishes, it is nevertheless a simple analytic solution in two-derivative $D=4$ gravity with a genuine curvature singularity. The reason for this is simply that the curvature singularity does not necessarily imply the divergence of the potential that controls the geodesic approach to a singularity.    

The examples we investigated showed that the thermal product formula-motivated asymptotic analysis \cite{Dodelson:2025jff} in the mixed $(t,\mathbf{k})$ space may reveal a union of singularities that survive the Fourier transform to the standard position space $(t,\mathbf{x})$ and another infinite set of {\em phantom singularities} which are eliminated by the integration over $\mathbf{k}$. What precisely is the situation for other, more complicated black holes remains an open, but important question. Moreover, given the recently revealed infinite sets of constraints on the spectra of QNMs by the {\em spectral duality relation} \cite{Grozdanov:2024wgo,Grozdanov:2025ner,Grozdanov:2025ulc} and resulting novel QNM sum rules \cite{Grozdanov:2024wgo,Grozdanov:2025ner,Dodelson:2023vrw}, all related to the validity of the thermal product formula hypothesis \cite{Dodelson:2023vrw}, it is a natural question to ask whether these structures reveal any interesting singularity-related properties of black holes. We intend to return to these questions in the near future. 

Another natural direction for future research is to examine the consequences of the bouncing geodesics for cases beyond the standard gauge/gravity duality. These include asymptotically flat and asymptotically de Sitter spacetimes. Despite the absence of known and controlled dual theories in these cases, Theorem~\ref{t.hadam} still guarantees that bouncing geodesics are reflected in the analytic structure of the bulk correlators. Finally, it would also be interesting to use this framework in the context of cosmology and study the singularities that appear in models of the Big Bang.


\section*{Acknowledgements}

We would like to thank M.\ Bajec, N.\ Čeplak, C.\ Esper, I.\ Gusev, C.\ Iossa, H.\ Liu, R.\ Karlsson, M.\ Kulaxizi, V.\ Movrin, A.\ Parnachev, and G.\ Policastro for useful discussions. We also thank the organisers and other participants of the IGAP workshop on ``Black Hole Perturbations and Holography''. The work of S.G.\ is supported by the STFC Ernest Rutherford Fellowship ST/T00388X/1. The work is also supported by the research programme P1-0402 and the project J7-60121 of Slovenian Research Agency (ARIS). S.V.\ is supported by the Marie Skłodowska-Curie Actions programme GA-101177446 and the Slovenian Research and Innovation Agency (ARIS), contract number 5110-18/2025-5. M.V.\ is funded by the research programme P1-0402 and the project J7-60121 of Slovenian Research Agency (ARIS).


\appendix

\section{Examples of the Hadamard form}\label{a.exampleBB}

In this appendix, we demonstrate examples of the validity of the Hadamard theorem \ref{t.hadam}, and the different types of scalings found in even and odd dimensions. We do this considering two propagators in flat Minkowski spacetime, and the retarded Green's function in $D=3$ anti-de Sitter space.

\subsection{Flat Minkowski space}

We first demonstrate the Hadamard form in an even number of spacetime
dimensions using the example of a flat Minkowski spacetime in $D=4$. In this case, the retarded Green's function of a massless scalar field is given by (see e.g.~Ref.~\cite{friedlander1975wave})
\begin{equation}
\CG (\{t,\mathbf{x}\},\{t',\mathbf{x}'\})
=
\frac{\theta(t-t')}{2\pi}\,\delta(\CL^2),
\end{equation}
where
\begin{equation}
\CL^2 = -(t-t')^2 + |\mathbf{x}-\mathbf{x}'|^2 .
\end{equation}
Thus, the Green's function is supported only on the lightcone $\CL^2=0$, in agreement with the Hadamard form from Theorem~\ref{t.hadam} in Eq.~\eqref{e.had}.

As noted below Theorem~\ref{t.hadam}, we assume that $D \geq 3$. To see the relevance of this assumption, one only needs to consider the massless retarded Green's function in $D=2$, where it is given by 
\begin{equation}
\CG (\{t,x\},\{t',x'\})
=
\frac{1}{2}\,\theta(t-t')\,\theta(-\CL^2),
\end{equation}
where $\CL^2 = -(t-t')^2+(x-x')^2$. Thus, in $D=2$, the retarded Green's function has support throughout the
interior of the future lightcone, rather than only on $\CL^2=0$.

Finally, we also state the result for odd number of spacetime dimensions, i.e., in $D=3$. There, the retarded correlator is given by
\begin{equation}
\CG(\{t,\mathbf{x}\},\{t',\mathbf{x}'\})
=
\frac{\theta(t-t')}{2\pi\sqrt{-\CL^2}}\,\theta(-\CL^2),
\end{equation}
where, again, we find agreement with the standard Hadamard forms stated in Eq.~\eqref{e.had}.

These flat-space examples illustrate the local structure underlying the
Hadamard theorem, whose proof is {\em local} on Lorentzian manifolds and is therefore governed by the Minkowski behaviour in a neighbourhood of any
point.

\subsection{Anti-de Sitter spacetime with \texorpdfstring{$D=3$}{D=3}}

Next, we consider a more complicated case of the empty AdS$_3$,
\begin{equation}\label{e.simpleST}
    \dd s^2=\frac{1}{z}(-\dd t^2+\dd z^2+\dd x^2),
\end{equation}
where $z=\frac1r$ and the conformal boundary is located at $z=0$. 

\subsection*{The geodesic computation:}  
To simplify the geodesic computations, set $x=0$ and assume a general space-like geodesic from point $\{z'=\frac{1}{r_i},\,t'\}$ to point $\{z=\frac{1}{r_f},\,t\}$. To compute the proper length, we can use Eq.~\eqref{e.lenghtgen} with $f(r)=r^2$,
\begin{equation}\label{e.EXpropl}
    \mathcal{L}=-\int_{r_i}^{r_f}\frac{\dd r}{\sqrt{E^2+r^2}}=-\ln(r_f+\sqrt{E^2+r_f^2})+\ln(r_i+\sqrt{E^2+r_i}).
\end{equation}
In a similar fashion one finds
\begin{equation}\label{e.EXtime}
    t-t'=-\int_{r_i}^{r_f}\frac{E\dd r}{r^2\sqrt{E^2+r^2}}=-\frac{\sqrt{E^2+r_f^2}}{Er_f}+\frac{\sqrt{E^2+r_i^2}}{Er_i}=\sqrt{z+\frac{1}{E^2}}-\sqrt{z'+\frac{1}{E^2}}.
\end{equation}
Inverting \eqref{e.EXtime} we get $E$ as a function of the initial and final positions; plugging this into Eq.~\eqref{e.EXpropl} we get the final expression for proper length written purely in terms of the coordinates $t$, $t'$, $z$ and $z'$ as
\begin{equation}\label{e.EXfingeo}
    \mathcal{L}(t,z,t',z')=\ln \left(\frac{\sqrt{\frac{4 (t-t')^2}{(t-t')^4-2 (t-t')^2
   \left(z^2+z'^2\right)+\left(z^2-z'^2\right)^2}+\frac{1}{z'^2}}+\frac{1}{z'}}{\sqrt{\frac{4 (t-t')^2}{(t-t')^4-2 (t-t')^2
   \left(z^2+z'^2\right)+\left(z^2-z'^2\right)^2}+\frac{1}{z^2}}+\frac{1}{z}}\right) .
\end{equation}
Now we fix $z'$, $t$ and $t'$ and treat $\mathcal{L}$ as a function of $z$. Define
\begin{equation}\label{e.primitdef}
    z_\pm\equiv z'\pm(t-t') ,
\end{equation}
which are the values for which the geodesic probe approaches the speed of light, i.e., the null limit of the space-like geodesic. Indeed, expanding \eqref{e.EXfingeo} in the $z\to z_+$ we find
\begin{equation}
    \mathcal{L}(z\to z_+)=\frac{\sqrt{2}\sqrt{t-t'}}{\sqrt{z'(t-t'+z')}}\sqrt{z_+-z}+\mathcal{O}(z_+-z)\longrightarrow0 .
\end{equation}

According to the Theorem \ref{t.hadam} this means that in the $z\to z_+$ limit the bulk-to-bulk retarded Green's function $\mathcal{G}$ should diverge as 
\begin{equation}\label{e.prediction}
    \mathcal{G}(z\to z_+)\sim\frac{1}{\mathcal{L}}\sim\frac{1}{\sqrt{z_+-z}} .
\end{equation}
Let us now compare this prediction with the actual Green's function.

\subsection*{Comparison:} 

For the spacetime in \eqref{e.simpleST}, the concrete form of the bulk-to-bulk retarded Green's function is known analytically (see e.g.~\cite{Balasubramanian:2012tu}):
\begin{equation}\label{e.prvka}
    \mathcal{G}(\{t,z,x\},\{t',z',x'\})=C\theta(t-t')\Im\left[\sigma^{-\Delta}{}_2F_1\Big(\frac{\Delta+1}{2},\frac{\Delta}{2};\Delta;\sigma^{-2}\Big)\right] ,
\end{equation}
where $C$ is a normalization constant and $\sigma$ is defined by
\begin{equation}
    \sigma\equiv\frac12\frac{z'^2+z^2-(t-t')^2+(x-x')^2+i\epsilon}{zz'}=\frac12\frac{(z-z_+)(z-z_-)}{zz'}+1 ,
\end{equation}
where the $\epsilon$ prescription (that we will from now on omit) determines from which sides one approaches the cuts, and in the second equality we defined
\begin{equation}
    z_\pm\equiv z'\pm T,\qq{where}T^2=(t-t')^2-(x-x')^2 ,
\end{equation}
in accordance with \eqref{e.primitdef}. 

To simplify Eq.~\eqref{e.prvka} we use the following identity:
\begin{equation}\label{e.ID1}
    {}_2F_1(a,b;a+b-\frac12;x)=\frac{2^{a+b-\frac32}}{\sqrt{1-x}}\Gamma(a+b-\frac12)x^{\frac{3-2a-2b}{4}}P_{-a+b-\frac12}^{-a-b+\frac32}(\sqrt{1-x}) ,
\end{equation}
where $P^\mu_\nu(y)$ is the associated Legendre polynomial, for which the following identity will be useful:
\begin{equation}\label{e.ID2}
    P_0^\mu(y)=\frac{1}{\Gamma(1-\mu)}\frac{(1+y)^\frac{\mu}{2}}{(1-y)^{\frac{\mu}{2}}} .
\end{equation}
Using Eqs.\ \eqref{e.ID1} and \eqref{e.ID2}, we rewrite the retarded Green's function as
\begin{equation}
    \mathcal{G}(\{t,z,x\},\{t',z',x'\})=C\theta(t-t')\Im\left[\sigma^{-1}\frac{2^{\Delta-1}}{\sqrt{1-\sigma^{-2}}}\left(\frac{1-\sqrt{1-\sigma^{-2}}}{1+\sqrt{1-\sigma^{-2}}}\right)^{\frac{\Delta-1}{2}}\right] .
\end{equation}
We are interested in the case where the two points become light-like separated. Following the same logic as in the geodesic computation, we fix all positions except $z$ and take the limit $z\to z_+$, in which case $\sigma\to1$. In this limit, the retarded bulk-to-bulk Green's function indeed diverges due to the factor $1/\sqrt{1-\sigma^{-2}}$. Concretely, we find
\begin{equation}
    \mathcal{G}(z\to z_+)\approx -C\theta(t-t')\frac{z_+z'}{\sqrt{z_+-z_-}}\frac{1}{\sqrt{z_+-z}},
\end{equation}
which is also in agreement with the prediction from the Hadamard theorem \eqref{e.prediction}.


\section{Thermal correlators and their analytic continuation}\label{a.spectraldensity}
In this appendix, we outline some basic definitions and properties of thermal correlators (see e.g.\ Refs.\ \cite{Bellac:2011kqa,Cuniberti_2001}) and discuss their geometric interpretation in the bulk. 

\subsection{Definitions and analytic continuations}

Let $\mathcal{O}(t,\mathbf{x})$ be a hermitian scalar operator, and let the thermal expectation value be
\begin{equation}
    \expval{\ldots}_\beta\equiv \frac1Z\Tr\left\{e^{-\beta H}\ldots \right\}.
\end{equation}
The two-sided correlator $G_{12}(x,x')$ of $\mathcal{O}(t,\mathbf{x})$ is defined as
\begin{equation}
    G_{12}(x)=\expval{\mathcal{O}\qty(t-i\frac{\beta}{2},\mathbf{x})\mathcal{O}(0)}_\beta.
\end{equation}
It is analytic on the strip $-\beta/2<\Im t<\beta/2$ and may be analytically extended beyond. The spectral function is defined as
\begin{equation}
    \rho(x,x')=\frac{1}{2}\expval{\left[\mathcal{O}(t,\mathbf{x}),\mathcal{O}(t',\mathbf{x}')\right]}_\beta,
\end{equation}
and may be expressed with the two-sided correlator as
\begin{equation}
    \rho(x)=\frac{1}{2}\lim_{\epsilon \rightarrow 0^+}\qty[G_{12}\qty(t+i\frac{\beta}{2}-i\epsilon,\mathbf{x})-G_{12}\qty(t-i\frac{\beta}{2}+i\epsilon,\mathbf{x})].\label{appeq:rho-twosided}
\end{equation}
Finally, the retarded Green's function $G(x)$ is defined as
\begin{equation}
    G(x)=2i\theta(t) \rho(x). \label{appeq:retarded}
\end{equation}
In Fourier space, we have
\begin{align}
    G_{12}(\omega,\mathbf k)&=\frac{\rho(\omega,\mathbf{k})}{\sinh \frac{\beta\omega}{2}},\\    \rho(\omega,\mathbf{k})&=\frac{G(\omega,\mathbf{k})-G(-\omega,\mathbf{k})}{2i}.
\end{align}
The $i\epsilon$ prescriptions in Eq.~\eqref{appeq:rho-twosided} assure that, in a causal theory, the spectral function vanishes outside the lightcone, and that the retarded Green's function vanishes outside the future lightcone, i.e.:
\begin{align}
    \rho(x)&= \theta(t^2-\mathbf{x}^2) \tilde \rho(x),\\
    G(x)&= 2i\theta(t)\theta(t^2-\mathbf{x}^2) \tilde \rho(x).
\end{align}
The Heaviside step function does not have a natural analytic extension, and may be analytically extended separately for its negative or positive argument. To understand the analytical extension of $G(x)$ for complex $t$, we therefore pick a time $t$ inside the future lightcone, and analytically extend from there. In other words, in the scope of this work, we are looking at the analytic extension of $\tilde \rho(x)$.

\subsection{Geometric interpretation}

\begin{figure}[ht!]
\centering
\includegraphics[scale=1.20]{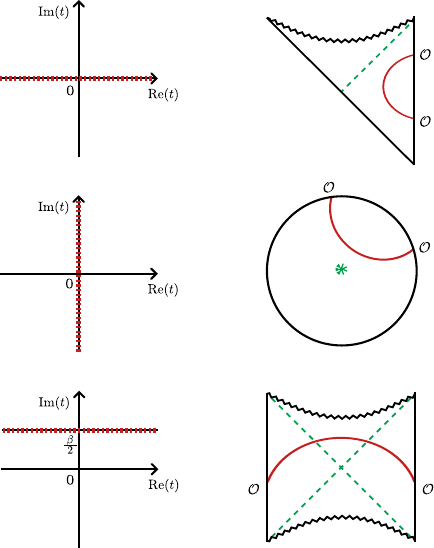}
\caption{Correlator domains in the complex time plane (shown by red dashed lines in the left column) and their geometric interpretations in the bulk (right column). From top to bottom, these are: 1-sided Lorentzian picture, 1-sided Euclidean picture and 2-sided Lorentzian picture for the AdS black brane. Red solid lines are sketches of time-like (top) and space-like (bottom) geodesics, and green dashed curves represent the horizon.}
\label{f.analko}
\end{figure}

Let us now discuss the bulk interpretation of the analytically continued correlators. For real temporal and spatial displacement one obtains the boundary retarded (resp., advanced) correlator by solving the bulk equation of motion with ingoing (resp., outgoing) boundary condition at the black hole horizon, Dirichlet condition at the conformal boundary and by applying the standard holographic dictionary. For this setup, the proper length $\mathcal{L}$ that enters the Theorems \ref{t.hadam} and \ref{t.bdryhadam} has a clear geometric interpretation as a real or complex\footnote{I.e., both the coordinates $\{r,t,\mathbf{x}\}$ and the turning point $r_T$ can be complex along the geodesic, while still connecting the chosen (real) anchoring points. As a relevant example, see, e.g., Ref.~\cite{Araya:2026shz}.} geodesic connecting the two operators on the boundary; see the top diagram in Fig.\ \ref{f.analko}. This is the so-called ``1-sided Lorentzian picture''.

Analytically continuing the correlator to purely imaginary time is equivalent to Wick rotating the metric into a Euclidean circle, see the middle diagram in Fig.\ \ref{f.analko} -- this is the so-called ``1-sided Euclidean picture''. In this case $\mathcal{L}$ is computed as a real (or complex) Euclidean geodesic in such spacetime.

Besides these two setups, the geometric interpretation of the analytically continued correlator exists for the situation where
\begin{equation}
    t=t_L+i\frac{\beta}{2},\qquad t_L\in\mathbb{R}.
\end{equation}
This setup can be interpreted as having two copies of a CFT on the two conformal boundaries in the maximally extended spacetime \cite{Maldacena:2001kr},\footnote{In this approach, the 1-sided thermal description is recovered by tracing over the Hilbert space of the second boundary CFT as in the thermofield double formalism for thermal field theories.} giving the so-called ``2-sided Lorentzian picture''. In this case, computing $\mathcal{L}$ that is relevant for the retarded propagator $G(t_L+i\frac{\beta}{2})$ is done by finding the geodesics that connect the operators on the two opposite boundaries, as depicted by a red curve in the bottom diagram of Fig.\ \ref{f.analko}. Such geodesics will naturally obtain an imaginary time-shift $i\frac{\beta}{2}$ as the two horizons are crossed in the complexified Schwarzschild coordinates.

Let us finish by making the following three remarks. First, note that assuming the 2-sided picture, we are secretly still computing the 1-sided setup but for operators separated by an additional constant imaginary factor. The singularities due to null geodesics (e.g., the bouncing geodesics) in this setup probe the \textit{same} retarded boundary propagator as was originally defined in the 1-sided picture, differing just by a simple analytic continuation. Both 1-sided and 2-sided pictures should therefore be thought of as investigating different time windows for the \textit{same} black hole and its dual correlator.

As a second remark, we note that for a generic point in the complex time (or position) space, the dual geometric interpretation is subtle, and to compute $\mathcal{L}$ one needs to assume complex geodesics with fully complex anchoring points. Nevertheless, we expect the Theorems \ref{t.hadam} and \ref{t.bdryhadam} to still hold.

The final remark concerns situations for which there exist multiple real or complex geodesics connecting the two boundary operators. As mentioned in the Sec.\ \ref{ss.HADthms}, in these situations, the Hadamard form is not guaranteed despite the fact that the singularity in the correlator is still expected to be present. The issues may appear when some of the geodesics become null, while others remain finite. This happens for the AdS black brane, where for the times between $0$ and a finite critical time $t_*$, there is one real geodesic (that in the $E\to\infty$ limit becomes null), and two complex geodesics (that remain non-zero in the $E\to\infty$ limit). As was shown in \cite{Afkhami-Jeddi:2025wra}, despite this subtlety, one still finds the singularity in the dual retarded correlator at the corresponding position, providing further evidence for the robustness of the Hadamard theorem~\ref{t.hadam} and Theorem~\ref{t.bdryhadam}.


\section{Fourier integrals}\label{a.fourik}
In this appendix, we provide the derivations of the Fourier integrals that allow us to analyse the complex $t$ behaviour of Green's function in the BTZ and the linear axion case. In both cases, we first focus on the two-sided correlator, since it is analytic for real times and therefore does not require a careful treatment with regards to integrating around various branch points. 

It will be useful to define
\begin{equation}
    f_\Delta(w)\equiv \Gamma\qty(\frac{\Delta}{2}\pm iw)
\end{equation}
for $\Re \Delta>0$. We use the notation by which
\begin{equation}
    \Gamma(a\pm b)=\Gamma(a+b)\Gamma(a-b).
\end{equation}
Its Fourier transform can be computed explicitly, e.g., using the beta function, and amounts to
\begin{equation}
    \int_{-\infty}^\infty \dd w \,f_\Delta(w)e^{-i\xi w}\equiv\frac{\pi \Gamma(\Delta)}{2^{\Delta-1}}\frac{1}{\cosh^\Delta \frac{\xi}{2}}.
\end{equation}

\subsection{The BTZ black hole with a scalar field}
\label{app:btz}
The $(t,x)$ representation of a scalar thermal correlator is completely determined by conformal invariance and the scaling dimension \cite{DiFrancesco:1997nk}. For the sake of completeness, we show how to compute the $(t,x)$ correlation functions from the $(\omega,k)$ representation of the correlator of the scalar primaries dual to a scalar field on the BTZ background. We start with the two-sided correlator in $(\omega,k)$ space
\begin{equation}
    \hat G_{12}(\omega,k)=\alpha\, \Gamma\qty(\frac{\Delta}{2}\pm i \frac{\beta}{4\pi}(\omega\pm k)),
\end{equation}
where $\alpha$ is an arbitrary normalisation constant. We are interested in
\begin{equation}
    G_{12}(t,x)=\frac{1}{(2\pi)^2}\int_{-\infty}^\infty \dd\omega \int_{-\infty}^\infty \dd k \, \hat G_{12}(\omega,k)\, e^{-i\omega t+ i k x}.
\end{equation}
By introducing the null coordinates
\begin{equation}
    u=\frac{\beta}{4\pi}(\omega+p), \qquad v=\frac{\beta}{4\pi}(\omega-p),
\end{equation}
the integral factorises,
\begin{equation}
    G_{12}(t,x)=\frac{2\alpha}{\beta^2} \int_{-\infty}^\infty \dd u \, f_\Delta(u) \,e^{-\frac{2\pi i}{\beta}u(t-x)}\int_{-\infty}^\infty \dd v \, f_\Delta(v)\, e^{-\frac{2\pi i}{\beta}v(t+x)},
\end{equation}
which gives
\begin{equation}
    G_{12}(t,x)=\frac{2\alpha}{\beta^2}\qty(\frac{\pi \Gamma(\Delta)}{2^{\Delta-1}})^2\frac{1}{\qty[\cosh\qty(\frac{\pi}{\beta}(t-x))\cosh\qty(\frac{\pi}{\beta}(t+x))]^\Delta}.
\end{equation}
The retarded correlator is recovered through Eqs.\ \eqref{appeq:rho-twosided} and \eqref{appeq:retarded}, and gives
\begin{equation}
    G(t,x)=\theta(t)\theta(t^2-x^2)\frac{4^{2-\Delta}\alpha \pi^2 \Gamma(\Delta)^2}{\beta^2}\frac{\sin\pi \Delta}{\qty[\sinh\qty(\frac{\pi}{\beta}(t-x))\sinh\qty(\frac{\pi}{\beta}(t+x))]^\Delta}.
\end{equation}
\subsection{The self-dual linear axion model}
\label{app:axion}
The two-sided correlator of the scalar primary in the case of the self-dual axion can be written as
\begin{equation}
    \hat G_{12}(\omega,k)=\alpha \,\Gamma\qty(\frac{5}{4}\pm i\frac{\beta}{4\pi}(\omega\pm p(k))),
\end{equation}
where
\begin{equation}
p(k)=\sqrt{p_0^2+k_x^2+k_y^2}, \quad p_0=\frac{\sqrt{7}\pi}{\beta},
\end{equation}
and $\alpha$ some arbitrary normalisation constant. We are interested in
\begin{equation}
    I(t)\equiv G_{12}(t,x=y=0)=\frac{1}{(2\pi)^3}\int_{-\infty}^\infty \dd\omega \int_{-\infty}^\infty \dd k_x \int_{-\infty}^\infty \dd k_y \,\hat G_{12}(\omega,k)\,e^{-i\omega t}.
\end{equation}
Introducing new coordinates $p$ and $\varphi$ as
\begin{equation}
    k_x=\sqrt{p^2-p_0^2}\cos \varphi, \qquad k_y=\sqrt{p^2-p_0^2}\sin\varphi,
\end{equation}
we get
\begin{equation}
    I(t)=\frac{1}{(2\pi)^2}\int_{-\infty}^\infty \dd \omega \int_{p_0}^\infty \dd p\, p \, \hat G_{12}(\omega,k(p))\, e^{-i\omega t}. 
\end{equation}
By using the fact that the integrand is odd in $p$, we can extend the integral over $p$ over the entire real axis by including the appropriate Heaviside functions
\begin{equation}
    \int_{p_0}^\infty \dd p \, p \, \hat G(\omega,k(p))=\frac{1}{2}\int_{-\infty}^\infty \dd p\, \qty[\theta(p-p_0)-\theta(-p-p_0)]\, p \,\hat G_{12}(\omega,k(p)).
\end{equation}
By inserting the Fourier representation of the Heaviside function
\begin{equation}
    \theta(p)=\frac{i}{2\pi}\int_{-\infty}^\infty \dd z \frac{e^{-ipz}}{z+i0^+},
\end{equation}
we get
\begin{equation}
I(t)=\frac{1}{(2\pi)^3}\int_{-\infty}^\infty \dd z \frac{e^{ip_0 z}}{z+i0^+}\int_{-\infty}^\infty \dd\omega \int_{-\infty}^\infty \dd p\, \hat G_{12}(\omega,k(p)) \, p \, \sin(pz)\, e^{-i\omega t}.
\end{equation}
We can now note that
\begin{equation}
     p \sin(pz)= -\frac{\partial}{\partial z} \cos(pz)
\end{equation}
and write
\begin{equation}
    I(t)=-\frac{1}{2\pi}\int_{-\infty}^\infty \dd z\frac{e^{ip_0 z}}{z+i0^+} \frac{\partial}{\partial z}W(t,z),
\end{equation}
where we have defined
\begin{equation}
    W(t,z)=\frac{1}{(2\pi)^2}\int_{-\infty}^\infty \dd \omega \int_{-\infty}^\infty \dd p \,\hat G_{12}(\omega,k(p))\, e^{-i\omega t-ipz},
\end{equation}
and used the fact that $\hat G(\omega,k(p))$ is even in $p$.
We can now move into `null' coordinates
\begin{equation}
    u=\frac{\beta}{4\pi}(\omega+p), \qquad v=\frac{\beta}{4\pi}(\omega-p),
\end{equation}
in which $\hat G(\omega,k(p))$ factorises as
\begin{equation}
    \hat G(\omega,k(p))=\alpha \,f_\frac{5}{2}(u)f_{\frac{5}{2}}(v),
\end{equation}
We then get
\begin{equation}
    W(t,z)= \frac{2\alpha}{\beta^2}\int_{\infty}^\infty \dd u \,f_{\frac{5}{2}}(u)\, e^{-\frac{2\pi i}{\beta}u(t+z)} \int_{\infty}^\infty \dd v\,f_{\frac{5}{2}}(v)\, e^{-\frac{2\pi i}{\beta}v(t-z)},
\end{equation}
which gives
\begin{equation}
    W(t,z)=\frac{9\alpha\pi^3}{64\beta^2}\frac{1}{\qty[\cosh\qty(\frac{\pi}{\beta}(t+z))\cosh\qty(\frac{\pi}{\beta}(t-z))]^\frac{5}{2}}.
\end{equation}
Finally, we get
\begin{equation}
    I(t)=\frac{45 \alpha \pi^3}{256\beta^3}\int_{-\infty}^\infty \dd z\frac{e^{ip_0 z}}{z+i0^+}\frac{\sinh \frac{2\pi z}{\beta}}{\qty[\cosh\qty(\frac{\pi}{\beta}(t+z))\cosh\qty(\frac{\pi}{\beta}(t-z))]^{\frac{7}{2}}}.
\end{equation}
Using the Sokhotski-Plemelj theorem, and defining $s=2\pi z/\beta$, we finally get
\begin{equation}
    I(t)=\frac{45\alpha \pi^3}{16\sqrt{2}\beta^3}\int_{-\infty}^\infty \dd s \frac{\sinh s}{s}\frac{\cos\frac{\sqrt{7}}{2}s}{\qty[\cosh\frac{2\pi t}{\beta}+\cosh{s}]^{\frac{7}{2}}}.
\end{equation}
In the case of the retarded Green's function, we would have to make appropriate choices regarding the integration contour, which would give rise to the appropriate Heaviside step functions.


\bibliographystyle{JHEP}
\bibliography{draft} 

\providecommand{\href}[2]{#2}\begingroup\raggedright\begin{thebibliography}{10}

\bibitem{Maldacena:1997re}
J.~M. Maldacena, {\it {The Large N limit of superconformal field theories and supergravity}},  {\em Adv. Theor. Math. Phys.} {\bf 2} (1998) 231--252, [\href{http://arxiv.org/abs/hep-th/9711200}{{\tt hep-th/9711200}}].

\bibitem{Gubser:1998bc}
S.~S. Gubser, I.~R. Klebanov, and A.~M. Polyakov, {\it {Gauge theory correlators from noncritical string theory}},  {\em Phys. Lett. B} {\bf 428} (1998) 105--114, [\href{http://arxiv.org/abs/hep-th/9802109}{{\tt hep-th/9802109}}].

\bibitem{Witten:1998qj}
E.~Witten, {\it {Anti-de Sitter space and holography}},  {\em Adv. Theor. Math. Phys.} {\bf 2} (1998) 253--291, [\href{http://arxiv.org/abs/hep-th/9802150}{{\tt hep-th/9802150}}].

\bibitem{Fidkowski:2003nf}
L.~Fidkowski, V.~Hubeny, M.~Kleban, and S.~Shenker, {\it {The Black hole singularity in AdS / CFT}},  {\em JHEP} {\bf 02} (2004) 014, [\href{http://arxiv.org/abs/hep-th/0306170}{{\tt hep-th/0306170}}].

\bibitem{Festuccia:2005pi}
G.~Festuccia and H.~Liu, {\it {Excursions beyond the horizon: Black hole singularities in Yang-Mills theories. I.}},  {\em JHEP} {\bf 04} (2006) 044, [\href{http://arxiv.org/abs/hep-th/0506202}{{\tt hep-th/0506202}}].

\bibitem{Ceplak:2024bja}
N.~\v{C}eplak, H.~Liu, A.~Parnachev, and S.~Valach, {\it {Black Hole Singularity from OPE}},  {\em JHEP} {\bf 10} (2024) 105, [\href{http://arxiv.org/abs/2404.17286}{{\tt arXiv:2404.17286}}].

\bibitem{Valach:2025saf}
S.~Valach, {\em {Thermal Correlators and Black Holes: From Infinity to Singularity}}.
\newblock PhD thesis, Trinity College Dublin. School of Mathematics. Discipline of Pure {\&} Applied Mathematics, TCD, Dublin, 2025.
\newblock \href{http://arxiv.org/abs/2508.17139}{{\tt arXiv:2508.17139}}.

\bibitem{Ceplak:2025dds}
N.~{\v{C}}eplak, H.~Liu, A.~Parnachev, and S.~Valach, {\it {Fooling the Censor: Going beyond inner horizons with the OPE}},  \href{http://arxiv.org/abs/2511.09638}{{\tt arXiv:2511.09638}}.

\bibitem{Araya:2026shz}
I.~J. Araya, C.~Esper, Y.~Jia, M.~Kulaxizi, and A.~Parnachev, {\it {Bulkcone Singularities and Complex Geodesics}},  \href{http://arxiv.org/abs/2602.12893}{{\tt arXiv:2602.12893}}.

\bibitem{Afkhami-Jeddi:2025wra}
N.~Afkhami-Jeddi, S.~Caron-Huot, J.~Chakravarty, and A.~Maloney, {\it {Imprint of the black hole singularity on thermal two-point functions}},  \href{http://arxiv.org/abs/2510.21673}{{\tt arXiv:2510.21673}}.

\bibitem{Jia:2025jbi}
H.~F. Jia and M.~Rangamani, {\it {Thermal spectral function asymptotics and black hole singularity in holography}},  \href{http://arxiv.org/abs/2512.15114}{{\tt arXiv:2512.15114}}.

\bibitem{AliAhmad:2026wem}
S.~Ali~Ahmad, A.~Almheiri, and S.~Lin, {\it {Continuing past the inner horizon using WKB}},  \href{http://arxiv.org/abs/2601.02354}{{\tt arXiv:2601.02354}}.

\bibitem{Jia:2026pmv}
Y.~Jia and M.~Kulaxizi, {\it {Bulk Phase Shift and Singularity}},  \href{http://arxiv.org/abs/2602.06558}{{\tt arXiv:2602.06558}}.

\bibitem{Dodelson:2025jff}
M.~Dodelson, C.~Iossa, and R.~Karlsson, {\it {Bouncing off a stringy singularity}},  \href{http://arxiv.org/abs/2511.09616}{{\tt arXiv:2511.09616}}.

\bibitem{hadamard1923lectures}
J.~Hadamard, {\em Lectures on Cauchy's problem in linear partial differential equations}.
\newblock Dover, 1923.

\bibitem{friedlander1975wave}
F.~G. Friedlander, {\em The wave equation on a curved space-time}, vol.~2.
\newblock Cambridge university press, 1975.

\bibitem{Dodelson:2023nnr}
M.~Dodelson, C.~Iossa, R.~Karlsson, A.~Lupsasca, and A.~Zhiboedov, {\it {Black hole bulk-cone singularities}},  {\em JHEP} {\bf 07} (2024) 046, [\href{http://arxiv.org/abs/2310.15236}{{\tt arXiv:2310.15236}}].

\bibitem{Dodelson:2023vrw}
M.~Dodelson, C.~Iossa, R.~Karlsson, and A.~Zhiboedov, {\it {A thermal product formula}},  {\em JHEP} {\bf 01} (2024) 036, [\href{http://arxiv.org/abs/2304.12339}{{\tt arXiv:2304.12339}}].

\bibitem{garabedian1998partial}
P.~Garabedian, {\em Partial Differential Equations}.
\newblock AMS Chelsea Publishing Series. AMS Chelsea, 1998.

\bibitem{Kay:1996hj}
B.~S. Kay, M.~J. Radzikowski, and R.~M. Wald, {\it {Quantum field theory on space-times with a compactly generated Cauchy horizon}},  {\em Commun. Math. Phys.} {\bf 183} (1997) 533--556, [\href{http://arxiv.org/abs/gr-qc/9603012}{{\tt gr-qc/9603012}}].

\bibitem{ikawa2000hyperbolic}
M.~Ikawa, {\em Hyperbolic partial differential equations and wave phenomena}, vol.~2.
\newblock American Mathematical Soc., 2000.

\bibitem{futurepaper}
S.~Grozdanov, V.~Movrin, and S.~Valach, {\it {Bouncing Geodesics, Singularities, and the Cavity Thermal Product Formula in Asymptotically Flat and de Sitter Black Holes}},  \href{http://arxiv.org/abs/2606.11297}{{\tt arXiv:2606.11297}}.

\bibitem{Hubeny:2006yu}
V.~E. Hubeny, H.~Liu, and M.~Rangamani, {\it {Bulk-cone singularities {\&} signatures of horizon formation in AdS/CFT}},  {\em JHEP} {\bf 01} (2007) 009, [\href{http://arxiv.org/abs/hep-th/0610041}{{\tt hep-th/0610041}}].

\bibitem{Dodelson:2020lal}
M.~Dodelson and H.~Ooguri, {\it {Singularities of thermal correlators at strong coupling}},  {\em Phys. Rev. D} {\bf 103} (2021), no.~6 066018, [\href{http://arxiv.org/abs/2010.09734}{{\tt arXiv:2010.09734}}].

\bibitem{Auzzi:2025sep}
R.~Auzzi, S.~Baiguera, L.~Guo, G.~Nardelli, and N.~Zenoni, {\it {Probing the bubble interior with entanglement entropy and bulk-cone singularities}},  {\em JHEP} {\bf 02} (2026) 215, [\href{http://arxiv.org/abs/2509.21632}{{\tt arXiv:2509.21632}}].

\bibitem{Buric:2025anb}
I.~Buri{\'c}, I.~Gusev, and A.~Parnachev, {\it {Thermal holographic correlators and KMS condition}},  {\em JHEP} {\bf 09} (2025) 053, [\href{http://arxiv.org/abs/2505.10277}{{\tt arXiv:2505.10277}}].

\bibitem{Buric:2025fye}
I.~Buri{\'c}, I.~Gusev, and A.~Parnachev, {\it {Holographic Correlators from Thermal Bootstrap}},  \href{http://arxiv.org/abs/2508.08373}{{\tt arXiv:2508.08373}}.

\bibitem{Barrat:2025twb}
J.~Barrat, D.~N. Bozkurt, E.~Marchetto, A.~Miscioscia, and E.~Pomoni, {\it {Analytic thermal bootstrap meets holography}},  \href{http://arxiv.org/abs/2510.20894}{{\tt arXiv:2510.20894}}.

\bibitem{Giombi:2026kdz}
S.~Giombi, Y.-Z. Li, and J.~Shan, {\it {Bouncing singularities and thermal correlators on line defects}},  \href{http://arxiv.org/abs/2603.11012}{{\tt arXiv:2603.11012}}.

\bibitem{Faruk:2023uzs}
M.~M. Faruk, E.~Morvan, and J.~P. van~der Schaar, {\it {Static sphere observers and geodesics in Schwarzschild-de Sitter spacetime}},  {\em JCAP} {\bf 05} (2024) 118, [\href{http://arxiv.org/abs/2312.06878}{{\tt arXiv:2312.06878}}].

\bibitem{Faruk:2025bed}
M.~M. Faruk, F.~Rost, and J.~P. van~der Schaar, {\it {Quasinormal modes and the switchback effect in Schwarzschild-de Sitter}},  {\em JHEP} {\bf 07} (2025) 050, [\href{http://arxiv.org/abs/2501.01388}{{\tt arXiv:2501.01388}}].

\bibitem{futurepaper2}
S.~Grozdanov, V.~Movrin, and S.~Valach, {\it {Quasinormal modes as exterior probes of black hole interiors in our Universe}},  \href{http://arxiv.org/abs/2608.13643}{{\tt arXiv:2608.13643}}.

\bibitem{NejcesAxion}
N.~\v{C}eplak, {\it {Complex bounces and black hole singularities}},  to appear.

\bibitem{Fitzpatrick:2019zqz}
A.~L. Fitzpatrick and K.-W. Huang, {\it {Universal Lowest-Twist in CFTs from Holography}},  {\em JHEP} {\bf 08} (2019) 138, [\href{http://arxiv.org/abs/1903.05306}{{\tt arXiv:1903.05306}}].

\bibitem{Karlsson:2022osn}
R.~Karlsson, A.~Parnachev, V.~Prilepina, and S.~Valach, {\it {Thermal stress tensor correlators, OPE and holography}},  {\em JHEP} {\bf 09} (2022) 234, [\href{http://arxiv.org/abs/2206.05544}{{\tt arXiv:2206.05544}}].

\bibitem{Huang:2022vet}
K.-W. Huang, R.~Karlsson, A.~Parnachev, and S.~Valach, {\it {Freedom near lightcone and ANEC saturation}},  {\em JHEP} {\bf 05} (2023) 065, [\href{http://arxiv.org/abs/2210.16274}{{\tt arXiv:2210.16274}}].

\bibitem{Esper:2023jeq}
C.~Esper, K.-W. Huang, R.~Karlsson, A.~Parnachev, and S.~Valach, {\it {Thermal stress tensor correlators near lightcone and holography}},  {\em JHEP} {\bf 11} (2023) 107, [\href{http://arxiv.org/abs/2306.00787}{{\tt arXiv:2306.00787}}].

\bibitem{Amado:2008hw}
I.~Amado and C.~Hoyos-Badajoz, {\it {AdS black holes as reflecting cavities}},  {\em JHEP} {\bf 09} (2008) 118, [\href{http://arxiv.org/abs/0807.2337}{{\tt arXiv:0807.2337}}].

\bibitem{Andrade:2013gsa}
T.~Andrade and B.~Withers, {\it {A simple holographic model of momentum relaxation}},  {\em JHEP} {\bf 05} (2014) 101, [\href{http://arxiv.org/abs/1311.5157}{{\tt arXiv:1311.5157}}].

\bibitem{Davison:2014lua}
R.~A. Davison and B.~Gout{\'e}raux, {\it {Momentum dissipation and effective theories of coherent and incoherent transport}},  {\em JHEP} {\bf 01} (2015) 039, [\href{http://arxiv.org/abs/1411.1062}{{\tt arXiv:1411.1062}}].

\bibitem{Grozdanov:2019uhi}
S.~Grozdanov, P.~K. Kovtun, A.~O. Starinets, and P.~Tadi{\'c}, {\it {The complex life of hydrodynamic modes}},  {\em JHEP} {\bf 11} (2019) 097, [\href{http://arxiv.org/abs/1904.12862}{{\tt arXiv:1904.12862}}].

\bibitem{Grozdanov:2025ner}
S.~Grozdanov and M.~Vrbica, {\it {Duality and four-dimensional black holes: Gravitational waves, algebraically special solutions, pole skipping, and the spectral duality relation in holographic thermal CFTs}},  {\em Phys. Rev. D} {\bf 112} (2025), no.~6 066019, [\href{http://arxiv.org/abs/2505.14229}{{\tt arXiv:2505.14229}}].

\bibitem{Natario:2004jd}
J.~Natario and R.~Schiappa, {\it {On the classification of asymptotic quasinormal frequencies for d-dimensional black holes and quantum gravity}},  {\em Adv. Theor. Math. Phys.} {\bf 8} (2004), no.~6 1001--1131, [\href{http://arxiv.org/abs/hep-th/0411267}{{\tt hep-th/0411267}}].

\bibitem{Grozdanov:2016vgg}
S.~Grozdanov, N.~Kaplis, and A.~O. Starinets, {\it {From strong to weak coupling in holographic models of thermalization}},  {\em JHEP} {\bf 07} (2016) 151, [\href{http://arxiv.org/abs/1605.02173}{{\tt arXiv:1605.02173}}].

\bibitem{Grozdanov:2018gfx}
S.~Grozdanov and A.~O. Starinets, {\it {Adding new branches to the {\textquotedblleft}Christmas tree{\textquotedblright} of the quasinormal spectrum of black branes}},  {\em JHEP} {\bf 04} (2019) 080, [\href{http://arxiv.org/abs/1812.09288}{{\tt arXiv:1812.09288}}].

\bibitem{Jansen:2017oag}
A.~Jansen, {\it {Overdamped modes in Schwarzschild-de Sitter and a Mathematica package for the numerical computation of quasinormal modes}},  {\em Eur. Phys. J. Plus} {\bf 132} (2017), no.~12 546, [\href{http://arxiv.org/abs/1709.09178}{{\tt arXiv:1709.09178}}].

\bibitem{Grozdanov:2025ulc}
S.~Grozdanov and M.~Vrbica, {\it {Thermal field theory correlators in the large-N limit and the spectral duality relation}},  {\em JHEP} {\bf 02} (2026) 106, [\href{http://arxiv.org/abs/2509.18074}{{\tt arXiv:2509.18074}}].

\bibitem{festucciathesis}
G.~Festuccia, {\em Black hole singularities in the framework of gauge/string duality}.
\newblock PhD thesis, Massachusetts Institute of Technology, 2007.

\bibitem{Festuccia:2008zx}
G.~Festuccia and H.~Liu, {\it {A Bohr-Sommerfeld quantization formula for quasinormal frequencies of AdS black holes}},  {\em Adv. Sci. Lett.} {\bf 2} (2009) 221--235, [\href{http://arxiv.org/abs/0811.1033}{{\tt arXiv:0811.1033}}].

\bibitem{erdelyi}
A.~Erd{\'e}lyi, W.~Magnus, F.~Oberhettinger, and F.~G. Tricomi, {\em Higher Transcendental Functions, Vol. 1}.
\newblock Bateman Manuscript Project. McGraw-Hill, New York, 1953.

\bibitem{DiFrancesco:1997nk}
P.~Di~Francesco, P.~Mathieu, and D.~Senechal, {\em {Conformal Field Theory}}.
\newblock Graduate Texts in Contemporary Physics. Springer-Verlag, New York, 1997.

\bibitem{Son:2002sd}
D.~T. Son and A.~O. Starinets, {\it {Minkowski space correlators in AdS / CFT correspondence: Recipe and applications}},  {\em JHEP} {\bf 09} (2002) 042, [\href{http://arxiv.org/abs/hep-th/0205051}{{\tt hep-th/0205051}}].

\bibitem{Grozdanov:2024wgo}
S.~Grozdanov and M.~Vrbica, {\it {Duality Constraints on Thermal Spectra of 3D Conformal Field Theories and 4D Quasinormal Modes}},  {\em Phys. Rev. Lett.} {\bf 133} (2024), no.~21 211601, [\href{http://arxiv.org/abs/2406.19790}{{\tt arXiv:2406.19790}}].

\bibitem{Balasubramanian:2012tu}
V.~Balasubramanian, A.~Bernamonti, B.~Craps, V.~Ker{\"a}nen, E.~Keski-Vakkuri, B.~M{\"u}ller, L.~Thorlacius, and J.~Vanhoof, {\it {Thermalization of the spectral function in strongly coupled two dimensional conformal field theories}},  {\em JHEP} {\bf 04} (2013) 069, [\href{http://arxiv.org/abs/1212.6066}{{\tt arXiv:1212.6066}}].

\bibitem{Bellac:2011kqa}
M.~L. Bellac, {\em {Thermal Field Theory}}.
\newblock Cambridge Monographs on Mathematical Physics. Cambridge University Press, 3, 2011.

\bibitem{Cuniberti_2001}
G.~Cuniberti, E.~De~Micheli, and G.~A. Viano, {\it {Reconstructing the thermal Green functions at real times from those at imaginary times}},  {\em Commun. Math. Phys.} {\bf 216} (2001) 59--83, [\href{http://arxiv.org/abs/cond-mat/0109175}{{\tt cond-mat/0109175}}].

\bibitem{Maldacena:2001kr}
J.~M. Maldacena, {\it {Eternal black holes in anti-de Sitter}},  {\em JHEP} {\bf 04} (2003) 021, [\href{http://arxiv.org/abs/hep-th/0106112}{{\tt hep-th/0106112}}].

\end{thebibliography}\endgroup

\end{document}